\documentclass[journal,final,doublecolumn]{IEEEtran}
\usepackage{subfigure}
\usepackage{graphicx}
\usepackage{amsmath}
\usepackage{amssymb}
\usepackage{mathrsfs}
\usepackage{cite}
\usepackage{multirow}
\usepackage{caption}
\renewcommand{\IEEEQED}{\IEEEQEDopen}
\DeclareMathOperator{\tr}{Tr}
\DeclareMathOperator{\supp}{supp}
\DeclareMathOperator{\pr}{Pr}

\newtheorem{theorem}{Theorem}[section]
\newtheorem{lemma}[theorem]{Lemma}

\newtheorem{proposition}[theorem]{Proposition}
\newtheorem{assumption}[theorem]{Assumption}
\newtheorem{rmk}[theorem]{Remark}
\newtheorem{model}[theorem]{Model}
\newtheorem{problem}[theorem]{Problem}

\newtheorem{result}[theorem]{Result}

\usepackage{savesym}
\savesymbol{AND}
\usepackage{algorithmic}
\usepackage{algorithm}
\begin{document}
\title{A particle filter approach to approximate posterior Cram\'er-Rao lower bound$^\star$}
\author{Aditya~Tulsyan, Biao~Huang, R. Bhushan~Gopaluni and~J. Fraser~Forbes
\thanks{$^\star$A condensed version of this article has been published in: Tulsyan, A., Huang, B., Gopaluni, R.B., Forbes, J.F. ``A particle filter approach to approximate posterior Cram\'er-Rao lower bound: The case of hidden states". {IEEE Transactions on Aerospace and Electronic Systems}, vol 49, no. 4, 2013, In press.}
\thanks{Authors' addresses: A. Tulsyan, B. Huang and J.F. Forbes are with the Computer Process Control Group, Department of Chemical and Materials Engineering, University of Alberta, Edmonton T6G-2G6, Alberta, Canada, (e-mail: \{tulsyan; biao.huang; fraser.forbes\}@ualberta.ca); and R.B. Gopaluni is with the Process Modeling and Control Lab, Department of Chemical and Biological Engineering, University of
British Columbia, Vancouver V6T-1Z3, BC, Canada, (e-mail: bhushan.gopaluni@ubc.ca).}}
\maketitle
\begin{abstract}
The posterior Cram\'er-Rao lower bound (PCRLB) derived in \cite{T1998} provides a bound on the mean square error (MSE) obtained with any non-linear state filter. Computing the PCRLB involves solving complex, multi-dimensional expectations, which do not lend themselves to an easy analytical solution. Furthermore, any attempt to approximate it using numerical or simulation based approaches require a priori access to the true states, which may not be available, except in simulations or in carefully designed experiments. To allow recursive approximation of the PCRLB when the states are hidden or unmeasured, a new approach based on sequential Monte-Carlo (SMC) or particle filters (PF) is proposed. The approach uses SMC methods to estimate the hidden states using a sequence of the available sensor measurements. The developed method is general and can be used to approximate the PCRLB in non-linear systems with non-Gaussian state and sensor noise. The efficacy of the developed method is illustrated on two simulation examples, including a practical problem of ballistic target tracking at re-entry phase.
\end{abstract}
\section{Introduction}
\label{sec:Chap2S1}
Non-linear filtering is one of the most important Bayesian inferencing methods, with several key applications in: navigation \cite{Gus2002}, guidance \cite{GSE1995}, tracking \cite{C1984}, fault detection \cite{RD2004} and fault diagnosis \cite{N2004}. Within the Bayesian framework, a filtering problem aims at constructing a posterior filter density \cite{D2001}. 

In the last few decades, several tractable algorithms based on analytical and statistical approximation of the Bayesian filtering (e.g., extended Kalman filter (EKF) and unscented Kalman filter (UKF)) have been developed to allow tracking in non-linear SSMs \cite{A2002}. Although filters, such as EKF and UKF are efficient in tracking, their performance is often limited or affected by various numerical and statistical approximations. Despite the great practical interest in evaluating  the non-linear filters, it still remains one of the most complex problems in estimation theory \cite{S2001}. 

The Cram\'er-Rao lower bound (CRLB) defined as an inverse of the Fisher information matrix (FIM) provides a theoretical lower bound on the second-order error (MSE) obtained with any maximum-likelihood (ML) based unbiased state or parameter estimator. An analogous extension of CRLB to the class of Bayesian estimators was derived by \cite{V1968}, which is commonly referred to as the PCRLB. The PCRLB is defined as the inverse of the posterior Fisher information matrix (PFIM)  and provides a lower bound on the  MSE obtained with any non-linear filter \cite{T1998}. A full statistical characterization of any non-Gaussian posterior density requires all higher-order moments \cite{R2004}. As a result, the PCRLB does not fully characterize the accuracy of non-linear filters. Nonetheless, it is an important tool, as it only depends on: system dynamics; prior density of the states; and system noise characteristics \cite{NB2001}. 

The PCRLB has been widely used as a benchmark for: (i) assessing the quality of different non-linear filters; (ii) comparing performances of non-linear filters against that of an optimal filter; and (iii) determining whether the filter performance requirements are practical or not. Some of the key practical applications of the PCRLB include: comparison of several non-linear filters for ballistic target tracking \cite{ARB2002}; terrain navigation \cite{B1999}; and design of systems with pre-specified performance bounds \cite{NH2000}. The PCRLB is also widely used in several other areas related to: multi-sensor resource deployment (e.g., radar resource allocation \cite{GS2011}, sonobuoy deployment in submarine tracking  \cite{HB2004}); sensor positioning \cite{FSK2006}; and optimal observer trajectory for bearings-only tracking \cite{PD1998,HM1993}.

The original PCRLB formulation in \cite{V1968} is based on batch data, which often renders its computation impractical for multi-dimensional non-linear SSMs. Alternatively, a recursive version of the PCRLB was proposed by \cite{B1975} for scalar non-linear SSMs with additive Gaussian noise. Its extension to deal with multi-dimensional case was developed much later in \cite{G1980,D1995}, where the authors compared the information matrix of a non-linear SSM with that of a suitable linear system with Gaussian noise. In the seminal paper \cite{T1998}, the authors proposed an elegant approach to recursively compute the PCRLB for discrete-time,  non-linear SSMs. Compared to \cite{G1980,D1995}, the PCRLB formulation in \cite{T1998} is more general as it is applicable to multi-dimensional non-linear SSMs with non-Gaussian state and sensor noise.  An overview of the historical developments of the PCRLB, along with other critical discussions can be found in \cite{K1989}. 

The PCRLB in \cite{T1998} provides a recursive procedure to compute the lower bound for tracking in general non-linear SSMs, operating with the probability of detection ${\pr_d=1}$ and the probability of false alarm ${\pr_{f}=0}$. Since then, several modified versions of the PCRLB have also appeared, which allow tracking in situations, such as: measurement origin uncertainty (${\pr_d=1}$ and ${\pr_{f}\geq 0}$) \cite{HMG2002}; missed detection (${\pr_d\leq 1}$ and ${\pr_{f}=0}$) \cite{FRT2002}; and cluttered environments (${\pr_d\leq 1}$ and ${\pr_{f}\geq 0}$) \cite{HFR2006}. However, unlike the bound formulation given in \cite{T1998}, the modified versions of the lower bound are mostly for a special class of non-linear SSMs with additive Gaussian state and sensor noise.

Notwithstanding a recursive procedure to compute the PCRLB in \cite{T1998}, obtaining a closed form solution to it is non-trivial. This is due to the involved complex, multi-dimensional expectations with respect to the states and measurements, which do not lend themselves to an easy analytical solution, except in linear systems \cite{NB2001}, where the Kalman filter (KF) provides an exact solution to the PCRLB.

Several attempts have been made in the past to address the aforementioned issues. First, several authors considered approximating the PCRLB for systems with:  (i) linear state dynamics with additive Gaussian noise and non-linear measurement model \cite{NB2001,H2008}; (ii) linear and non-linear SSMs with additive Gaussian state and sensor noise \cite{S2001,M2010}; and (iii) linear SSMs with unknown measurement uncertainty \cite{X2005}. The special sub-class of non-linear SSMs with additive Gaussian noise allows reduction of the complex, multi-dimensional expectations to a lower dimension, which are relatively easier to approximate.
\section{Motivation and contributions}
\label{sec:Chap2S2}
To obtain a reasonable approximation to the PCRLB for general non-linear SSMs, several authors have considered using simulation based techniques, such as the Monte Carlo (MC) method. Although a MC method makes the lower bound computations off-line, nevertheless, it is a popular approach, since for many real-time applications in tracking and navigation, the design, selection and performance evaluation of different filtering algorithms are mostly done a priori or off-line. Furthermore, availability of huge amount of historical test-data, makes MC method a viable option. An MC based bound approximation have appeared for several systems with: target generated measurements \cite{ARB2002,H2008}; measurement origin uncertainty \cite{HMG2002}; cluttered environments \cite{HFR2006,MHLW2009}; and Markovian models \cite{LS2010,Bessel2003}. Although MC methods can be effectively used to approximate the involved expectations, with respect to the states and measurements, it requires an ensemble of the true states and measurements. While the sensor readings may be available from the historical test-data, the true states may not be available, except in simulations or in carefully designed experiments \cite{Lei2011}.

To avoid having to use the true states, \cite{Lei2011} proposed an EKF and UKF based method to compute the PCRLB formulation in \cite{T1998}. To approximate the bound, \cite{Lei2011} first assumes the densities associated with the expectations to be Gaussian, and then uses an EKF and UKF to approximate the Gaussian densities using an estimate of the mean and covariance. Even though the method proposed in \cite{Lei2011} is fast, since it only works with the first two statistical moments, there are several performance and applicability related issues with this numerical approach, such as: (i) relies on the linearisation of the underlying non-linear dynamics around the state estimates, which not only results in additional numerical errors, but also introduces bias in the PCRLB approximation; (ii) the method is applicable only for non-linear SSMs with additive Gaussian state and sensor noise; (iii)  convergence of the numerical solution to the theoretical lower bound is not guaranteed; (iv) provides limited control for improving the quality of the resulting numerical solution; and (v) it involves long and tedious  calculations of the first two moments of the assumed Gaussian densities. 

Recently, \cite{ZNV2011} derived a conditional lower bound for general non-linear SSMs, and used an SMC based method to approximate it in absence of the true states. Unlike the unconditional PCRLB in \cite{T1998}, the conditional PCRLB can be computed in real-time; however, as shown in \cite{ZNV2011}, the bound in less optimistic (or higher) compared to the unconditional PCRLB. This limits its use to applications, where real-time bound computation is far more important than obtaining a tighter limit on the tracking performance. However, in applications, such as filter design and selection, where the primary focus is on devising an efficient filtering strategy, the PCRLB in \cite{T1998} provides an optimistic measure of the filter performance. 

To the authors' best knowledge, there are no known numerical method to approximate the unconditional PCRLB in \cite{T1998}, when the true states are unavailable.

The following are the main contributions in this paper: (i) an SMC based method is developed to numerically approximate the unconditional PCRLB in \cite{T1998}, for a general stochastic non-linear SSMs operating with ${\pr_d=1}$ and ${\pr_{f}=0}$. The expectations defined originally with respect to the true states and measurements are reformulated to accommodate use of the available sensor readings. This is done by first conditioning the distribution of the true states over the sensor readings, and then using an SMC method to approximate it. (ii) Based on the above developments, a numerical method to compute the lower bound for a class of discrete-time, non-linear SSMs with additive Gaussian state and sensor noise is derived. This is required, since several practical problems, especially in tracking, navigation and sensor management, are often modelled as non-linear SSMs, with additive Gaussian noise. (iii) Convergence results for the SMC based PCRLB approximation is also provided. (iii) The quality of the SMC based PCRLB approximation is illustrated on two examples, which include a uni-variate, non-stationary growth model and a practical problem of ballistic target tracking at re-entry phase.

The proposed simulation based method is an off-line method, which can be used to deliver an efficient numerical approximation to the lower bound in \cite{T1998}, based on the sensor readings alone. Compared to the EKF and UKF based PCRLB approximation method derived in \cite{Lei2011}, the proposed SMC based method: (i) is far more general as it can approximate the PCRLB for a larger class of discrete-time, non-linear SSMs with possibly non-Gaussian state and sensor noise; (ii) avoids numerical errors arising due to the use of dynamics linearisation methods; and (iii) provides a far greater control over the quality of the resulting approximation. Moreover, several theoretical results exist for the SMC methods, which can be used to suggest convergence of the SMC based PCRLB approximation to the actual lower bound. All these features of the proposed method are either validated theoretically or illustrated on simulation examples.
\section{Problem formulation}
\label{sec:Chap2S3}
In this paper, we consider a model for a class of general stochastic non-linear systems.
\begin{model}
\label{Chap2Model1}
Consider the following discrete-time, stochastic non-linear SSM
\begin{subequations}
 \label{eq:Chap2E1}
 \begin{align}
    {X}_{t+1}=&{f}_t({X}_{t},{u}_{t},{\theta}, {V}_{t}),	\label{eq:Chap2E1a}\\
    {Y}_t=&{g}_t({X}_{t},{u}_{t},{\theta}, {W}_{t}),		\label{eq:Chap2E1b}
 \end{align}
 \end{subequations}
where: ${{X}_t \in{\mathcal{X}}\subseteq \mathbb{R}^{n}}$ and ${{Y}_t\in{\mathcal{Y}}\subseteq\mathbb{R}^{m}}$ are the state variables and sensor measurements, respectively; ${{u}_{t}\in\mathcal{U}\subseteq\mathbb{R}^{p}}$ is input variables and ${{\theta}\in{\Theta}\subseteq\mathbb{R}^{r}}$ are the model parameters. Also: the state and sensor noise are represented as ${{V}_{t}\in \mathbb{R}^{n}}$ and ${{W}_{t}\in \mathbb{R}^{m}}$, respectively. ${f_t(\cdot)}$ is an n-dimensional state mapping function and ${g_t(\cdot)}$ is a m-dimensional measurement mapping function, where each being possibly non-linear in its arguments.
\end{model}
Model \ref{Chap2Model1} represents one of the most general classes of discrete-time, stochastic non-linear SSMs. For notational simplicity, explicit dependence on ${u_t\in\mathcal{U}}$ and ${\theta\in\Theta}$ are not shown in the rest of this article; however, all the derivations that appear in this paper hold with $u_t$ and $\theta$ included.  Assumptions on Model \ref{Chap2Model1} are discussed next.  
\begin{assumption}
\label{Chap2A1}
The state and sensor dynamics are defined as ${{f_t}:=\mathcal{X}\times \mathbb{R}^{n} \rightarrow \mathbb{R}^{n}}$ and ${{g_t}:=\mathcal{X}\times \mathbb{R}^{m}\rightarrow \mathbb{R}^{m}}$, respectively, are at least twice differentiable with respect to ${X_t\in\mathcal{X}}$. Also, the parameters ${\theta\in\Theta}$ and inputs ${u_t\in\mathcal{U}}$ are assumed to be known {a priori}.
\end{assumption}
\begin{assumption}
\label{Chap2A2E}
Sensor measurements are target-originated, operating with probability of false alarm ${\pr_f=0}$ and probability of detection ${\pr_d=1}$. The target states ${X_t\in\mathcal{X}}$ are hidden Markov process, observed only through the measurement process ${Y_t\in\mathcal{Y}}$.
\end{assumption}
\begin{assumption}
\label{Chap2A2}
${V}_{t}$, ${W}_{t}$ and $X_0$ are mutually independent sequences of independent random variables described by the probability density functions (pdfs) ${p}(v_t)$, $p(w_t)$ and $p(x_0)$, respectively. These pdfs are known in their classes (e.g., Gaussian; uniform) and are parametrized by a known and finite number of moments (e.g., mean; variance).
\end{assumption}
\begin{assumption}
\label{Chap2A3}
For a random realization $(x_{t+1},x_t,v_t)\in\mathcal{X}\times\mathcal{X}\times\mathbb{R}^n$ and $(y_t,x_t,w_t)\in\mathcal{Y}\times\mathcal{X}\times\mathbb{R}^m$ satisfying Model \ref{Chap2Model1}, $\nabla_{v_t}f^T_t(x_t,v_t)$ and $\nabla_{w_t}g^T_t(x_t,w_t)$ have rank $n$ and $m$, such that using implicit function theorem, ${p(x_{t+1}|x_t)=p(V_t=\tilde{f}_t(x_t,x_{t+1}))}$ and $p(y_{t}|x_t)=p(W_t=\tilde{g}_t(x_t,y_{t}))$ do not involve Dirac delta functions. 
\end{assumption}
\subsection{Posterior Cram\'{e}r-Rao lower bound}
\label{sec:Chap2S3.1}
The conventional CRLB provides a lower bound on the MSE of any ML based estimator. An analogous extension of the CRLB to the class of Bayesian estimators was derived by \cite{V1968}, and is referred to as the PCRLB inequality. Extension of the PCRLB to non-linear tracking was provided by \cite{T1998}, and is given next.
\begin{lemma}
\label{lemma:Chap2L1}
Let $\{Y_{1:t}\}_{t\in\mathbb{N}}$ be a sequence from Model \ref{Chap2Model1}, then MSE of any tracking filter at $t\in\mathbb{N}$ is bounded from below by the following matrix inequality
\begin{align}
P_{t|t}\triangleq\mathbb{E}_{p(X_{0:t},Y_{1:t})}[(X_t-\widehat{X}_{t|t})(X_t-\widehat{X}_{t|t})^T]\succcurlyeq J_t^{-1},\label{eq:Chap2E2}
\end{align}
where: $P_{t|t}$ is a ${n\times n}$ matrix of MSE; ${\widehat{X}_{t|t}\triangleq\widehat{X}_t(Y_{1:t}):=\mathbb{R}^{tm}\rightarrow \mathbb{R}^n}$ is a point estimate of ${X_t\in\mathcal{X}}$ at time ${t\in\mathbb{N}}$, given the measurement sequence ${\{{Y_{1:t}=y_{1:t}\}\triangleq\{y_1,\dots,y_t\}}}$; $J_t$ is a $n\times n$ PFIM matrix; $J^{-1}_t$ is a $n\times n$ PCRLB matrix; ${p(x_{0:t},y_{1:t})}$ is a joint probability density of the states and measurements up until time ${t\in\mathbb{N}}$; the superscript $(\cdot)^T$ is the transpose operation; and $\mathbb{E}_{p(\cdot)}[\cdot]$ is the expectation operator with respect to the pdf  $p(\cdot)$. 
\end{lemma}
\begin{IEEEproof}
See \cite{V1968} for a detailed proof. 
\end{IEEEproof}
Inequality (\ref{eq:Chap2E2}) implies that ${P_{t|t}-J_t^{-1}\succcurlyeq 0}$ is a positive semi-definite matrix for all ${\widehat{X}_{t|t}\in\mathbb{R}^n}$ and ${t\in\mathbb{N}}$. (\ref{eq:Chap2E2}) can also be written in terms of a scalar MSE (SMSE) as
\begin{align}
\label{eq:Chap2E3}
P^S_{t|t}\triangleq\mathbb{E}_{p(X_{0:t},Y_{1:t})}[\|X_t-\widehat{X}_{t|t}\|^2]\geq \tr[J_t^{-1}],
\end{align}
where $\tr[\cdot]$ is the trace operator, and $\|\cdot\|$ is a 2-norm.
\begin{lemma}
\label{lemma:Chap2L2}
For a system represented by Model \ref{Chap2Model1} and operating under Assumptions \ref{Chap2A1} through \ref{Chap2A3}, the PFIM in Lemma \ref{lemma:Chap2L1} can be recursively computed as \cite{T1998,S2001}
\begin{align}
\label{eq:Chap2E4}
J_{t+1}=D_t^{22}-[D_t^{12}]^T(J_t+D_t^{11})^{-1}D_t^{12},
\end{align}
where:
\begin{subequations}
\label{eq:Chap2E5}
\begin{align}
D_t^{11}=&\mathbb{E}_{p(X_{0:t+1},Y_{1:t+1})}[-\Delta_{X_t}^{X_t}\log~p(X_{t+1}|X_t)];\label{eq:Chap2E5b}\\
D_t^{12}=&\mathbb{E}_{p(X_{0:t+1},Y_{1:t+1})}[-\Delta_{X_t}^{X_{t+1}}\log~p(X_{t+1}|X_t)];\label{eq:Chap2E5c}\\
D_t^{22}=&\mathbb{E}_{p(X_{0:t+1},Y_{1:t+1})}[-\Delta_{X_{t+1}}^{X_{t+1}}\log~p(X_{t+1}|X_t)\nonumber\\
&-\Delta_{X_{t+1}}^{X_{t+1}}\log~p(Y_{t+1}|X_{t+1})];\label{eq:Chap2E5e}
\end{align}
\end{subequations}
and: $\Delta$ is a Laplacian operator such that ${\Delta^{Y}_{X}\triangleq\nabla_{X}\nabla_{Y}^{T}}$ with ${\nabla_{X}\triangleq\left[\frac{\partial{}}{\partial{X}}\right]}$ being a gradient operator, evaluated at the true states. Also, $J_0=\mathbb{E}_{p(X_{0})}[-\Delta_{X_0}^{X_0}\log~p(X_{0})]$.
\end{lemma}
\begin{IEEEproof}
See \cite{T1998} for a complete proof. 
\end{IEEEproof}
For Model \ref{Chap2Model1}, obtaining a closed-form solution to the PFIM or PCRLB is  non-trivial. This is due to the complex integrals involved in (\ref{eq:Chap2E5}), which do not lend themselves to an easy analytical solution. The main problem addressed in this paper is discussed next.
\begin{problem}
\label{problem:Chap2P1}
Compute a numerical solution to the PCRLB given in Lemma \ref{lemma:Chap2L1} for systems represented by Model \ref{Chap2Model1} and operating under Assumptions \ref{Chap2A1} through \ref{Chap2A3}. 
\end{problem}
Use of simulation based methods in addressing Problem \ref{problem:Chap2P1} is discussed next.
\section{Approximating PCRLB}
\label{sec:Chap2S4}
MC method is a popular approach, which can be used to approximate the PCRLB; however, as discussed in Section \ref{sec:Chap2S2}, MC method requires an ensemble of true states and sensor measurements. While sensor readings may be available from the historical test-data, the true states may not be available in practice. To allow the use of sensor readings in approximating the PCRLB, this paper reformulates the integrals in (\ref{eq:Chap2E5}) as given below.
\begin{proposition}
\label{proposition:Chap2P1}
The complex, multi-dimensional expectations in (\ref{eq:Chap2E5}), with respect to the density $p(x_{0:t+1},y_{1:t+1})$ can be reformulated, and written as follows:
\begin{subequations}
\label{eq:Chap2E6}
\begin{align}
I_t^{11}=&\mathbb{E}_{p(X_{0:t+1}|Y_{1:t+1})}[-\Delta_{X_t}^{X_t}\log~p(X_{t+1}|X_t)];						\label{eq:Chap2E6a}\\
I_t^{12}=&\mathbb{E}_{p(X_{0:t+1}|Y_{1:t+1})}[-\Delta_{X_t}^{X_{t+1}}\log~p(X_{t+1}|X_t)];					\label{eq:Chap2E6b}\\
I_t^{22,a}=&\mathbb{E}_{p(X_{0:t+1}|Y_{1:t+1})}[-\Delta_{X_{t+1}}^{X_{t+1}}\log~p(X_{t+1}|X_t)];		\label{eq:Chap2E6c}\\
I_t^{22,b}=&\mathbb{E}_{p(X_{0:t+1}|Y_{1:t+1})}[-\Delta_{X_{t+1}}^{X_{t+1}}\log~p(Y_{t+1}|X_{t+1})],	\label{eq:Chap2E6d}
\end{align}
\text{where:}
\begin{align}
D_t^{11}&=\mathbb{E}_{p(Y_{1:t+1})}[I_t^{11}];				\label{eq:Chap2E6e}\\
D_t^{12}&=\mathbb{E}_{p(Y_{1:t+1})}[I_t^{12}];				\label{eq:Chap2E6f}\\
D_t^{22}&=\mathbb{E}_{p(Y_{1:t+1})}[I_t^{22,a}+I_t^{22,b}].	\label{eq:Chap2E6g}
\end{align}
\end{subequations}
\end{proposition}
\begin{IEEEproof}
The proof is based on decomposition of the pdf $p(x_{0:t+1},y_{1:t+1})$ in (\ref{eq:Chap2E5}), using the probability condition ${p(x_{0:t+1},y_{1:t+1})=p(y_{1:t+1})p(x_{0:t+1}|y_{1:t+1})}$. 
\end{IEEEproof}
\begin{rmk}
\label{remark:Chap2R1}
In Proposition \ref{proposition:Chap2P1} the integrals are with respect to $p(y_{1:t+1})$ and $p(x_{0:t+1}|y_{1:t+1})$. The advantage of representing (\ref{eq:Chap2E5}) as (\ref{eq:Chap2E6}) is evident: using historical test-data, expectations with respect to $p(y_{1:t+1})$ can be approximated using MC, while that defined with respect to $p(x_{0:t+1}|y_{1:t+1})$ can be approximated using an SMC method. 
\end{rmk}
\subsection{SMC based PCRLB approximation}
\label{sec:Chap2S4.1}
It is not our aim here to review SMC methods in details, but to simply highlight their role in approximating the multi-dimensional integrals in Proposition \ref{proposition:Chap2P1}. For a detailed exposition on SMC methods, see  \cite{D2001,R2004}. The essential idea behind SMC methods is to generate a large set of random particles (samples) from the target pdf, with respect to which the integrals are defined. The target pdf of interest in Proposition \ref{proposition:Chap2P1} is ${p}(x_{0:t}|y_{1:t})$. Using SMC methods, the target distribution, defined as $p(dx_{0:t+1}|y_{1:t+1})\triangleq p(x_{0:t+1}|y_{1:t+1})dx_{0:t+1}$ can be approximated as given below.
\begin{align}
\tilde{p}(dx_{0:t+1}|y_{1:t+1})=\sum_{i=1}^N W_{0:t+1|t+1}^i\delta_{X^i_{0:t+1|t+1}}(dx_{0:t+1}),\label{eq:Chap2EQ2}
\end{align}
where: $\tilde{p}(dx_{0:t+1}|y_{1:t+1})$ is an $N$-particle SMC approximation of the target distribution $p(dx_{0:t+1}|y_{1:t+1})$ and $\{X^i_{0:t+1|t+1};~W_{0:t+1|t+1}^i\}_{i=1}^{N}$ are the $N$ pairs of particle realizations and their associated weights distributed according to $p(x_{0:t+1}|y_{1:t+1})$, such that $\sum_{i=1}^NW_{0:t|t}^i=1$. Using (\ref{eq:Chap2EQ2}), an SMC approximation of (\ref{eq:Chap2E6a}), for example, can be computed as
\begin{align}
\tilde{I}_t^{11}=\sum_{i=1}^N W^i_{0:t+1|t+1}[-\Delta_{X_t}^{X_t}\log~p(X^i_{t+1|t+1}|X^i_{t|t+1})].\label{eq:Chap2E7aa}
\end{align}
where $\tilde{I}_t^{11}$ is an SMC estimate of $I_t^{11}$ and the Laplacian is evaluated at $\{X^i_{t:t+1|t+1}\}_{i=1}^N$. 

The convergence of (\ref{eq:Chap2E7aa}) to (\ref{eq:Chap2E6a}) depends on (\ref{eq:Chap2EQ2}). Many sharp results on convergence of SMC methods are available (see \cite{CD2002} for a survey paper and \cite{DM2004} for a book length review). A selection of these results highlighting the difficulties in approximating $p(dx_{0:t}|y_{1:t})$ with an SMC method are presented below.
\begin{theorem}
\label{theorem:Chap2T1}
For any bounded test function ${\phi_t: \mathcal{X}^{t+1} \rightarrow \mathbb{R}}$, there exists ${C_{t,p}<\infty}$, such that for any $p>0$, ${N\geq1}$ and $t\geq1$, the following inequality holds
\begin{align}
\label{eq:Chap2E21}
\mathbb{E}\bigg[\bigg|\int_{\mathcal{X}^{t+1}} \phi_t(x_{0:t})\epsilon_t(dx_{0:t}|y_{1:t})\bigg|^p\bigg]^{\frac{1}{p}}\leq\frac{C_{t,p}\bar{\phi}_t}{N^{1/2}},
\end{align} 
where ${\epsilon_t(dx_{0:t}|y_{1:t})= \tilde{p}(dx_{0:t}|y_{1:t})-{p}(dx_{0:t}|y_{1:t})}$ is the $N$-particle approximation error, $\bar{\phi}_t=\sup_{x_{0:t}\in\mathcal{X}^{t+1}}|\phi_t(x_{0:t})|$, and the expectation is with respect to the particle realizations. 
\end{theorem}
\begin{IEEEproof}
See Theorem 2 in \cite{PDD2003} for a detailed proof.
\end{IEEEproof}
\begin{rmk}
\label{remark:Chap2R2}
The result in Theorem \ref{theorem:Chap2T1} is weak, since ${C_{t,p}\in\mathbb{R}}$ being a function of ${t\in\mathbb{N}}$, grows exponentially/polynomially with time \cite{Kantas2009}. To guarantee a fixed precision of the approximation in (\ref{eq:Chap2E7aa}), $N$ has to increase with $t$. The result in Theorem \ref{theorem:Chap2T1} is not surprising, since  (\ref{eq:Chap2EQ2}) requires sampling from the pdf $p(x_{0:t}|y_{1:t})$, whose dimension increases as $n(t+1)$. In literature Theorem \ref{theorem:Chap2T1} is referred to as the sample path degeneracy problem. This is a fundamental limitation of SMC methods; wherein, for ${N\in\mathbb{N}}$, the quality of the approximation of $p(dx_{0:t}|y_{1:t})$ deteriorates with time.
\end{rmk}
The motivation to use SMC methods to approximate the complex, multi-dimensional  integrals in Proposition \ref{proposition:Chap2P1} is based on the fact that encouraging results can be obtained under the exponential forgetting assumption on Model \ref{Chap2Model1}. Since ${{\theta} \in{\Theta}}$ is assumed to be known (see Assumption \ref{Chap2A1}), the forgetting property in Model \ref{Chap2Model1} holds. With the forgetting property, it is possible to establish results of the form given in the next theorem.
\begin{theorem}
\label{theorem:Chap2T2}
For an integer $L>0$, and any bounded test function $\phi_L: \mathcal{X}^{L} \rightarrow \mathbb{R}$, there exists $D_{L,p}<\infty$, such that for any $p>0$, $N\geq1$ and $t\geq1$, the following inequality holds  
\begin{align}
\label{eq:Chap2E22}
\mathbb{E}\bigg[\bigg|\int_{\mathcal{X}^{L}} \phi_L(x_{t-L+1:t})\epsilon_L(dx_{t-L+1:t}|y_{1:t})\bigg|^p\bigg]^{\frac{1}{p}}\leq\frac{D_{L,p}\bar{\phi}_L}{N^{1/2}},
\end{align} 
where $\epsilon_L(dx_{t-L+1:t}|y_{1:t})= \int_{\mathcal{X}^{t-L+1}}\epsilon_t(dx_{0:t}|y_{1:t})$. 
\end{theorem}
\begin{IEEEproof}
See Theorem 2 in \cite{PDD2003} for a detailed proof.
\end{IEEEproof}
\begin{rmk}
\label{remark:Chap2R3}
Since ${D_{L,p}\in\mathbb{R}}$ is independent of ${t\in\mathbb{N}}$, Theorem \ref{theorem:Chap2T2} suggests that an SMC based approximation of the most recent marginal posterior pdf $p(x_{t-L+1:t}|y_{1:t})$, over a fixed horizon $L>0$ does not result in the error accumulation. 
\end{rmk}
For our purposes, to make the SMC based PCRLB approximation effective, the dimension of the integrals in Proposition \ref{proposition:Chap2P1} needs to be reduced. An SMC based approximation of the PCRLB over a reduced dimensional state-space is discussed next.
\begin{lemma}
\label{lemma:Chap2L3}
For a system represented by Model \ref{Chap2Model1}, using the Markov property of the target states in Assumptions \ref{Chap2A2E}, Proposition \ref{proposition:Chap2P1} can be written as follows:
\begin{subequations}
\begin{align}
I_t^{11}=&\mathbb{E}_{p(X_{t:t+1}|Y_{1:t+1})}[-\Delta_{X_t}^{X_t}\log~p(X_{t+1}|X_t)];\label{eq:Chap2EQE1a}\\
I_t^{12}=&\mathbb{E}_{p(X_{t:t+1}|Y_{1:t+1})}[-\Delta_{X_t}^{X_{t+1}}\log~p(X_{t+1}|X_t)];\label{eq:Chap2EQE1b}\\
I_t^{22,a}=&\mathbb{E}_{p(X_{t:t+1}|Y_{1:t+1})}[-\Delta_{X_{t+1}}^{X_{t+1}}\log~p(X_{t+1}|X_t)];\label{eq:Chap2EQE1c}\\
I_t^{22,b}=&\mathbb{E}_{p(X_{t+1}|Y_{1:t+1})}[-\Delta_{X_{t+1}}^{X_{t+1}}\log~p(Y_{t+1}|X_{t+1})].\label{eq:Chap2EQE1d}
\end{align}
\end{subequations}
\end{lemma}
\begin{IEEEproof}
The proof is based on a straightforward use of the definition of expectation and  Markov property of Model \ref{Chap2Model1}. For example, the integrals in (\ref{eq:Chap2E6a}) can be written as
\begin{subequations}
\begin{align}
I_t^{11}=&\int_{\mathcal{X}^{t+2}}[-\Delta_{x_t}^{x_t}\log~p(x_{t+1}|x_t)]p(dx_{0:t+1}|y_{1:t+1}),\\
=&\int_{\mathcal{X}^2}[-\Delta_{x_t}^{x_t}\log~p(x_{t+1}|x_t)]p(dx_{t:t+1}|y_{1:t+1}),\\
=&\mathbb{E}_{p(X_{t:t+1}|Y_{1:t+1})}[-\Delta_{X_t}^{X_t}\log~p(X_{t+1}|X_t)],\label{eq:Chap2E9}
\end{align}
\end{subequations}
where ${p(dx_{0:t+1}|y_{1:t+1})\triangleq p(x_{0:t+1}|y_{1:t+1})dx_{0:t+1}}$, and in (\ref{eq:Chap2E9}), since the integrand is independent of ${x_{0:t-1}\in\mathcal{X}^{t}}$, it is marginalized out of the integral. Equations (\ref{eq:Chap2EQE1b}) through (\ref{eq:Chap2EQE1d}) can be derived based on similar arguments, which completes the proof.  
\end{IEEEproof}
\begin{rmk}
\label{remark:Chap2R4}
The dimension of the expectations in (\ref{eq:Chap2E6a}) through (\ref{eq:Chap2E6c}) reduces from $n(t+2)$ to $2n$; whereas, in (\ref{eq:Chap2E6d}), it reduces from $n(t+2)$ to $n$ for all ${t\in\mathbb{N}}$. Moreover, since expectations in Lemma \ref{lemma:Chap2L3} are with respect to $p(x_{t:t+1}|y_{1:t+1})$ and $p(x_{t+1}|y_{1:t+1})$, an SMC method can be effectively used with a finite number of particles (see Theorem \ref{theorem:Chap2T2}). 
\end{rmk}
\subsection{General non-linear SSMs} 
\label{sec:Chap2S4.2}
To approximate the multi-dimensional integrals in Lemma \ref{lemma:Chap2L3} for Model \ref{Chap2Model1}, a set of randomly generated samples from the target distribution $p(dx_{t:t+1}|y_{1:t+1})$ is required. First note that the target pdf $p(x_{t:t+1}|y_{1:t+1})$ can alternatively be written as given below.
\begin{lemma}
\label{lemma:Chap2L4}
The target pdf $p(x_{t:t+1}|y_{1:t+1})$, with respect to which the integrals in Lemma \ref{lemma:Chap2L3} are defined can be decomposed, and written as
\begin{align}
\label{eq:Chap2E10}
p(x_{t:t+1}|y_{1:t+1})=\frac{p(x_{t+1}|x_t)p(x_t|y_{1:t})p(x_{t+1}|y_{1:t+1})}{\int_{\mathcal{X}}p(x_{t+1}|x_{t})p(dx_{t}|y_{1:t})}.
\end{align}
\end{lemma}
\begin{IEEEproof}
First note that the target pdf $p(x_{t:t+1}|y_{1:t+1})$ can be written as
\begin{align}
\label{eq:Chap2EB1}
p(x_{t:t+1}|y_{1:t+1})=p(x_t|x_{t+1},y_{1:t}, y_{t+1})p(x_{t+1}|y_{1:t+1}).
\end{align}
From the Markov property of (\ref{eq:Chap2E1}), and from the Bayes' theorem, (\ref{eq:Chap2EB1}) can be written as
\begin{subequations}
\begin{align}
&p(x_{t:t+1}|y_{1:t+1})\nonumber\\
=&\frac{p(y_{t+1}|x_t,x_{t+1},y_{1:t})p(x_t|x_{t+1},y_{1:t})p(x_{t+1}|y_{1:t+1})}{p(y_{t+1}|x_{t+1},y_{1:t})},\label{eq:Chap2EB2a}\\
=&\frac{p(y_{t+1}|x_{t+1},y_{1:t})p(x_t|x_{t+1},y_{1:t})p(x_{t+1}|y_{1:t+1})}{p(y_{t+1}|x_{t+1},y_{1:t})}\label{eq:Chap2EB2b},\\
=&p(x_t|x_{t+1},y_{1:t})p(x_{t+1}|y_{1:t+1}).\label{eq:Chap2EB2c}
\end{align}
\end{subequations}
Applying Bayes' theorem again in (\ref{eq:Chap2EB2c}) yields
\begin{subequations}
\begin{align}
&p(x_{t:t+1}|y_{1:t+1})\nonumber\\
=&\frac{p(x_{t+1}|x_t,y_{1:t})p(x_t|y_{1:t})p(x_{t+1}|y_{1:t+1})}{p(x_{t+1}|y_{1:t})}\label{eq:Chap2EB3a},\\
=&\frac{p(x_{t+1}|x_t)p(x_t|y_{1:t})p(x_{t+1}|y_{1:t+1})}{\int_{\mathcal{X}}p(x_{t+1}|x_{t})p(dx_{t}|y_{1:t})},\label{eq:Chap2EB3b}
\end{align}
\end{subequations}
where in (\ref{eq:Chap2EB3b}), the Law of Total Probability is used, which completes the proof.
\end{IEEEproof}
\begin{rmk}
\label{remark:Chap2R5}
The procedure for generating random particles from densities, such as the uniform or Gaussian, is well described in literature; however, due to the multi-variate, and non-Gaussian nature of the target pdf, generating random particles from $p(x_{t:t+1}|y_{1:t+1})$ is a non-trivial problem. An alternative idea is to employ an importance sampling function (ISF), from which random particles are easier to generate \cite{D2001}. 
\end{rmk}
In this paper, the product of two pdfs in (\ref{eq:Chap2E10}) is selected as the ISF, such that
\begin{align}
\label{eq:Chap2EQ3}
q(x_{t:t+1}|y_{1:t+1})\triangleq p(x_t|y_{1:t})p(x_{t+1}|y_{1:t+1}),
\end{align}
where $q(x_{t:t+1}|y_{1:t+1})$ is a non-negative ISF on $\mathcal{X}^2$, such that   
$\supp q(x_{t:t+1}|y_{1:t+1})\supseteq\supp p(x_{t:t+1}|y_{1:t+1})$. Choice of an ISF similar to (\ref{eq:Chap2EQ3}) was also employed in \cite{T2001,S2011} to develop a particle smoothing algorithm for discrete-time, non-linear SSMs. Thus to be able to generate random samples from (\ref{eq:Chap2EQ3}), samples from the two posteriors $p(x_t|y_{1:t})$ and $p(x_{t+1}|y_{1:t+1})$ need to be generated first. Again, using the principles of ISF, particles from the posterior pdf can be generated using any advanced SMC methods (e.g., ASIR \cite{APF1999}, resample-move algorithm \cite{WRC2002}, block sampling strategy \cite{AMS2006}) or for example, using the method in \cite{S2011,G2008}. The method described in \cite{S2011,G2008} is outlined in Algorithm \ref{algorithm:Chap2A0}.
\newcounter{ALC@tempcntr}
\newcommand{\LCOMMENT}[1]{%
    \setcounter{ALC@tempcntr}{\arabic{ALC@rem}}
    \setcounter{ALC@rem}{1}
    \item 
    \setcounter{ALC@rem}{\arabic{ALC@tempcntr}}
}
\begin{algorithm}[!t]
  \caption{SMC based posterior density approximation}
  \label{algorithm:Chap2A0}
  \begin{algorithmic}[1]
  \LCOMMENT ~ \textbf{Input:} Given Model \ref{Chap2Model1}, satisfying Assumptions \ref{Chap2A1} through \ref{Chap2A3}, assume a prior pdf on $X_0$, such that ${X_0\sim p(x_0)}$. Also, select algorithm parameter $N$.
	\LCOMMENT ~ \textbf{Output:} Recursive SMC approximation of the posterior $p(dx_t|y_{1:t})$ for all $t\in\mathbb{N}$.
    \STATE Generate $N$ independent and identically distributed particles ${\{{X}^i_{0|-1}\}_{i=1}^N\sim p({x}_0)}$ and set the associated weights to $\{W^i_{0|-1}=N^{-1}\}_{i=1}^N$. Set $t\leftarrow 1$.
    \STATE Sample ${\{X^i_{t|t-1}\}_{i=1}^N\sim p(x_t|y_{1:t-1})}$. Set ${\{W^i_{t|t-1}=N^{-1}\}_{i=1}^N}$.
    \WHILE{$t\in\mathbb{N}$}
    \STATE Use ${\{Y_t=y_t\}}$ and compute the importance weights $\{W^i_{t|t}\}_{i=1}^N$ using
    \begin{align}
	\label{eq:Chap2E27}
	W^i_{t|t}= \frac{W^i_{t|t-1}p(y_t|X^i_{t|t-1})}{\sum_{j=1}^NW^j_{t|t-1}p(y_t|X^i_{t|t-1})}.
	\end{align}
    \STATE Resample the particle set $\{{X}^j_{t|t}\}_{j=1}^N$ with replacement from $\{X^i_{t|t-1}\}_{i=1}^N$, such that
\begin{align}
\label{eq:Chap2E28}
\pr({X}^j_{t|t}=X^i_{t|t-1})=W^i_{t|t},
\end{align}
where $Pr(\cdot)$ is a probability measure. Set $\{W^i_{t|t}=N^{-1} \}_{i=1}^N$.
    \STATE Sample ${\{X^i_{t+1|t}\}_{i=1}^N\sim p(x_{t+1}|y_{1:t})}$ using (\ref{eq:Chap2E33}). Set ${\{W^i_{t+1|t}=N^{-1}\}_{i=1}^N}$.
    \STATE Set $t\leftarrow t+1$.
    \ENDWHILE
  \end{algorithmic}
\end{algorithm}
It is important to note that in importance sampling, degeneracy is a common problem; wherein, after a few time instances, the density of the weights in (\ref{eq:Chap2E27}) become skewed. The resampling step in (\ref{eq:Chap2E27}) is crucial in limiting the effects of degeneracy. 
Finally using {Algorithm \ref{algorithm:Chap2A0}.}, the particle representation of $p(dx_t|y_{1:t})$ and $p(dx_{t+1}|y_{1:{t+1}})$ are given by
\begin{subequations}
\label{eq:Chap2E12}
\begin{align}
\tilde{p}(dx_t|y_{1:t})&=\frac{1}{N}\sum_{i=1}^N\delta_{{X}^i_{t|t}}(dx_t)\label{eq:Chap2E12a},\\
\tilde{p}(dx_{t+1}|y_{1:t+1})&=\frac{1}{N}\sum_{j=1}^N\delta_{{X}^j_{t+1|t+1}}(dx_{t+1}).\label{eq:Chap2E12b}
\end{align}
\end{subequations}
Here ${\{{X}^i_{t|t}\}_{i=1}^N\sim \tilde{p}(x_t|y_{1:t})}$ and ${\{{X}^i_{t+1|t+1}\}_{i=1}^N\sim \tilde{p}(x_{t+1}|y_{1:t+1})}$ are the $N$ pairs of resampled i.i.d.~samples from ${\tilde{p}(x_t|y_{1:t})}$ and ${\tilde{p}(x_{t+1}|y_{1:t+1})}$, respectively.
\begin{rmk}
\label{remark:Chap2R6}
Uniform convergence in time of (\ref{eq:Chap2E12}) has been established by  \cite{DM2004,NC2004}. Although these results rely on strong mixing assumptions of Model \ref{Chap2Model1}, uniform convergence has been observed in numerical studies for a wide class of non-linear time-series models, where the mixing assumptions are not satisfied.
\end{rmk}
Substituting (\ref{eq:Chap2E12}) into (\ref{eq:Chap2EQ3}), yields an SMC approximation of the ISF, i.e.,
\begin{align}
\tilde{q}(dx_{t:t+1}|y_{1:t+1})=&\frac{1}{N^2}\sum_{j=1}^N\sum_{i=1}^N\delta_{X^i_{t|t}, X^j_{t+1|t+1}}(dx_{t:t+1}),\label{eq:Chap2E13}
\end{align}
where $\tilde{q}(dx_{t:t+1}|y_{1:t+1})$ is an $N^2$-particle SMC approximation of the ISF distribution ${q}(dx_{t:t+1}|y_{1:t+1})$ and $\{X^i_{t|t};~X^j_{t+1|t+1}\}_{i=1,j=1}^{N,N}\sim \tilde{q}(x_{t:t+1}|y_{1:t+1})$ are particles from the ISF.
\begin{lemma}
\label{lemma:Chap2L5}
An SMC approximation of the target distribution ${p}(dx_{t:t+1}|y_{1:t+1})$ can be computed using the SMC approximation of ${q}(dx_{t:t+1}|y_{1:t+1})$ given in (\ref{eq:Chap2E13}), such that
\begin{align}
\label{eq:Chap2EQ5}
\tilde{p}&(dx_{t:t+1}|y_{1:t+1})=\sum_{i=1}^NW^{i}_{t|t, t+1|t+1}\delta_{{X}^{i}_{t|t},{X}^{i}_{t+1|t+1}}(dx_{t:t+1}),
\end{align}
where:
\begin{subequations}
\label{eq:Chap2E32}
\begin{align}
W^{i}_{t|t, t+1|t+1}&\triangleq\frac{\zeta^{i}_{t|t, t+1|t+1}}{\sum_{j=1}^N\zeta^{j}_{t|t, t+1|t+1}};\label{eq:Chap2E32a}\\
\zeta^{i}_{t|t, t+1|t+1}&\triangleq\frac{p({X}^i_{t+1|t+1}|{X}^i_{t|t})}{N\sum_{m=1}^Np({X}^i_{t+1|t+1}|{X}^m_{t|t})};\label{eq:Chap2E32b}
\end{align}
\end{subequations}
and $\tilde{p}(dx_{t:t+1}|y_{1:t+1})$ is an SMC approximation of the target distribution ${p}(dx_{t:t+1}|y_{1:t+1})$. 
\end{lemma}
\begin{IEEEproof}
Substituting (\ref{eq:Chap2E13}) into (\ref{eq:Chap2E10}) followed by several algebraic manipulations yields an SMC approximation of ${p}(dx_{t:t+1}|y_{1:t+1})$, denoted by $\tilde{p}(dx_{t:t+1}|y_{1:t+1})$, such that
\begin{subequations}
\label{eq:Chap2EC1}
\begin{align}
&\tilde{p}(dx_{t:t+1}|y_{1:t+1})\nonumber\\
=&\frac{p(x_{t+1}|x_t)\tilde{q}(dx_{t:t+1}|y_{1:t+1})}{\int_{\mathcal{X}}p(x_{t+1}|x_{t})\tilde{p}(dx_{t}|y_{1:t})},\label{eq:Chap2EC1a}\\
=&\frac{N p(x_{t+1}|x_t)\sum_{j=1}^N\sum_{i=1}^N\delta_{{X}^{i}_{t|t},{X}^{j}_{t+1|t+1}}(dx_{t:t+1})}{N^{2}\int_{\mathcal{X}}p(x_{t+1}|x_{t})\sum_{m=1}^N\delta_{{X}^i_{t|t}}(dx_t)},\label{eq:Chap2EC1b}
\end{align}
\begin{align}
=&\frac{\sum_{j=1}^N\sum_{i=1}^Np({X}^j_{t+1|t+1}|{X}^i_{t|t})\delta_{{X}^{i}_{t|t},{X}^{j}_{t+1|t+1}}(dx_{t:t+1})}{N\sum_{m=1}^Np({X}^j_{t+1|t+1}|{X}^m_{t|t})},\label{eq:Chap2EC1c}\\
=&\sum_{j=1}^N\sum_{i=1}^NW^{i,j}_{t|t, t+1|t+1}\delta_{{X}^{i}_{t|t},{X}^{j}_{t+1|t+1}}(dx_{t:t+1})\label{eq:Chap2EC1d},
\end{align}
\end{subequations}
where
\begin{align}
\label{eq:Chap2EC2}
W^{i,j}_{t|t, t+1|t+1}&\triangleq\frac{p({X}^j_{t+1|t+1}|{X}^i_{t|t})}{N\sum_{m=1}^Np({X}^j_{t+1|t+1}|{X}^m_{t|t})},
\end{align}
Equation (\ref{eq:Chap2EC1d}) is an SMC approximation of $p(dx_{t:t+1}|y_{1:t+1})$. The computational complexity of the weights in (\ref{eq:Chap2EC2}) is of the order $\mathcal{O}(N^2)$. As suggested in \cite{G2008}, without significant loss in the quality of the approximation, the complexity can be reduced to the order $\mathcal{O}(N)$ by replacing (\ref{eq:Chap2EC1d}) with (\ref{eq:Chap2EQ5}), which completes the proof. 
\end{IEEEproof}

The distribution of weights in (\ref{eq:Chap2EQ5}) becomes skewed after a few time instances. To avoid this, the particles in (\ref{eq:Chap2EQ5}) are resampled using systematic resampling, such that
\begin{align}
\label{eq:Chap2EQ7}
\pr({X}^{j}_{t:t+1|t+1}=\{{X}^{i}_{t|t};~{X}^{i}_{t+1|t+1}\})=W^{i}_{t|t,t+1|t+1},
\end{align}
where $\{{X}^{i}_{t:t+1|t+1}\}_{i=1}^N\sim \tilde{p}(x_{t:t+1}|y_{1:t+1})$ are resampled i.i.d. particles. With resampling, the SMC approximation of the target distribution in (\ref{eq:Chap2EQ5}) can be represented as
\begin{align}
\label{eq:Chap2EQ8}
\tilde{p}(dx_{t:t+1}|y_{1:t+1})=\frac{1}{N}\sum_{i=1}^N\delta_{X^i_{t:t+1|t+1}}(dx_{t:t+1}).
\end{align}
Expectation in Lemma \ref{lemma:Chap2L3}, with respect to the marginalized pdf $p(x_{t}|y_{1:t+1})$ (see (\ref{eq:Chap2EQE1d})) can also be approximated using SMC methods as given in the next lemma. 
\begin{lemma}
\label{lemma:Chap2L6}
Let $\{{X}^{i}_{t:t+1|t+1}\}_{i=1}^N$ in (\ref{eq:Chap2EQ8}) be i.i.d.~resampled particles distributed according to $\tilde{p}({x}_{t:t+1}| {y}_{1:t+1})$ then an SMC approximation of $p(dx_{t}|{y}_{1:t+1})$ is given by
\begin{align}
\label{eq:Chap2EL5}
\tilde p(dx_{t}|y_{1:t+1})=\frac{1}{N}\sum_{i=1}^{N} \delta_{X^i_{t|t+1}}(dx_t), 
\end{align}
where $\tilde{p}(dx_{t}|y_{1:t+1})$ is an SMC approximation of $p(dx_{t}|y_{1:t+1})$ and $\delta_{X^i_{t|t+1}}(\cdot)$ is a marginalized Dirac delta function in $dx_{t}$, centred around the random particle $X^i_{t|t+1}$.
\end{lemma}
\begin{IEEEproof}
See \cite{ABBF2012} for the proof.
\end{IEEEproof}

Lemma \ref{lemma:Chap2L6} gives a procedure for computing an SMC approximation of $p(dx_{t}|y_{1:t+1})$, using the particles from the SMC approximation of ${p}(d{x}_{t:t+1}| {y}_{1:t+1})$. Expectation with respect to  $p(y_{1:t+1})$ in  Proposition \ref{proposition:Chap2P1} can be approximated using MC method, such that
\begin{align}
\label{eq:Chap2EQ9}
\tilde{p}(dy_{1:t+1})=\frac{1}{M}\sum_{j=1}^M\delta_{Y^j_{1:t+1}}(dy_{1:t+1}),
\end{align}
where $\tilde{p}(dy_{1:t+1})$ is an MC approximation of ${p}(dy_{1:t+1})$, and $M$   is the total number of i.i.d. measurement sequences obtained from the historical test-data. Note that the approximation in (\ref{eq:Chap2EQ9}) is possible only under Assumption \ref{Chap2A1}; however, in general, estimating the marginalized likelihood function ${p}(y_{1:t+1})$ is non-trivial \cite{Kantas2009}.

Finally, an SMC approximation of the PCRLB for systems represented by Model \ref{Chap2Model1} and operating under Assumptions \ref{Chap2A1} through \ref{Chap2A3} is summarized in the next lemma.
\begin{lemma}
\label{lemma:Chap2L7}
Let a general stochastic non-linear system be represented by Model \ref{Chap2Model1}, such that it satisfies Assumption \ref{Chap2A1} through \ref{Chap2A3}. Let ${\{Y_{1:t}=y^j_{1:t}\}_{j=1}^M}$ be ${M\in\mathbb{N}}$ i.i.d.~measurement sequences generated from Model \ref{Chap2Model1}, then the matrices (\ref{eq:Chap2E5b}) through (\ref{eq:Chap2E5e}) in Lemma \ref{lemma:Chap2L2} can be recursively approximated as follows:
\begin{subequations}
\begin{align}
\tilde{D}_t^{11}=&-\frac{1}{MN}\sum_{j=1}^M\sum_{i=1}^N[\Delta_{X_t}^{X_t}\log~p(X^{i,j}_{t+1|t+1}|X^{i,j}_{t|t+1})];\label{eq:Chap2EE38a}\\
\tilde{D}_t^{12}=&-\frac{1}{MN}\sum_{j=1}^M\sum_{i=1}^N[\Delta_{X_t}^{X_{t+1}}\log~p(X^{i,j}_{t+1|t+1}|X^{i,j}_{t|t+1})];\label{eq:Chap2EE38b}\\
\tilde{D}_t^{22}=&-\frac{1}{MN}\sum_{j=1}^M\sum_{i=1}^N[\Delta_{X_{t+1}}^{X_{t+1}}\log~p(X^{i,j}_{t+1|t+1}|X^{i,j}_{t|t+1})\nonumber\\
&+\Delta_{X_{t+1}}^{X_{t+1}}\log~p(Y^j_{t+1}|X^{i,j}_{t+1|t+1})]; \label{eq:Chap2EE38c}
\end{align}
\end{subequations}
and ${\{X^{i,j}_{t:t+1|t+1}\}_{i=1}^{N}\sim p(x_{t:t+1}|y^j_{1:t+1})}$ is a set of $N$ resampled particles from (\ref{eq:Chap2EQ8}), distributed according to $p(x_{t:t+1}|y^j_{1:t+1})$ for all ${\{Y_{1:t+1}=y^j_{1:t+1}\}_{j=1}^M}$. 
\end{lemma}
\begin{IEEEproof}
For a measurement sequence ${\{Y_{1:t}=y^j_{1:t}\}}$, an SMC approximation of the target distribution in (\ref{eq:Chap2EQ8}) can be written as
\begin{align}
\label{eq:Chap2EQ10}
\tilde{p}(dx_{t:t+1}|y^j_{1:t+1})=\frac{1}{N}\sum_{i=1}^N\delta_{X^{i,j}_{t:t+1|t+1}}(dx_{t:t+1}),
\end{align}
where ${X^{i,j}_{t:t+1|t+1}\sim p(x_{t:t+1}|y^j_{1:t+1})}$ are resampled particles. Substituting (\ref{eq:Chap2EQ10}) into Lemma \ref{lemma:Chap2L3}, an SMC approximation of (\ref{eq:Chap2EQE1a})  through (\ref{eq:Chap2EQE1d}) can be obtained as follows:
\begin{subequations}
\label{eq:Chap2EQE2}
\begin{align}
\tilde{I}_t^{11}=&\frac{1}{N}\sum_{i=1}^N-\Delta_{X_t}^{X_t}\log~p(X^{i,j}_{t+1|t+1}|X^{i,j}_{t|t+1});\label{eq:Chap2EQE2a}\\
\tilde{I}_t^{12}=&\frac{1}{N}\sum_{i=1}^N-\Delta_{X_t}^{X_{t+1}}\log~p(X^{i,j}_{t+1|t+1}|X^{i,j}_{t|t+1});\label{eq:Chap2EQE2b}\\
\tilde{I}_t^{22,a}=&\frac{1}{N}\sum_{i=1}^N-\Delta_{X_{t+1}}^{X_{t+1}}\log~p(X^{i,j}_{t+1|t+1}|X^{i,j}_{t|t+1});\label{eq:Chap2EQE2c}\\
\tilde{I}_t^{22,b}=&\frac{1}{N}\sum_{i=1}^N-\Delta_{X_{t+1}}^{X_{t+1}}\log~p(Y^j_{t+1}|X^{i,j}_{t+1|t+1}),\label{eq:Chap2EQE2d}
\end{align}
where $\tilde{I}_t$ is an SMC approximation of ${I}_t$. Substituting (\ref{eq:Chap2EQE2}) and (\ref{eq:Chap2EQ9}) into (\ref{eq:Chap2E6e}) through (\ref{eq:Chap2E6g}) yields (\ref{eq:Chap2EE38a}) through (\ref{eq:Chap2EE38c}), which completes the proof.
\end{subequations}
\end{IEEEproof}
Lemma \ref{lemma:Chap2L7} gives an SMC based numerical method to approximate the complex, multi-dimensional integrals in Lemma \ref{lemma:Chap2L2}. Note that since Lemma \ref{lemma:Chap2L7} is valid for a general non-linear SSMs, the derivatives of the logarithms of the pdfs in (\ref{eq:Chap2EE38a}) through (\ref{eq:Chap2EE38c}) are left in its original form, but can be computed for a given system.

Building on the developments in this section, an SMC approximation of the PCRLB for a class of non-linear SSMs with additive Gaussian noise is presented next.
\subsection{Non-linear SSMs with additive Gaussian noise}
\label{sec:Chap2S4.3}
Many practical applications in tracking (e.g., ballistic target tracking \cite{ARB2002}, bearings-only tracking \cite{C1998}, range-only tracking \cite{Song1999}, multi-sensor resource deployment \cite{HB2004} and other navigation problems \cite{KGK2003}) can be described by non-linear SSMs with additive Gaussian noise. Since the class of practical problems with additive Gaussian noise is extensive, especially in tracking, navigation and sensor management, an SMC based numerical method for approximating the PCRLB for such class of non-linear systems is presented.
\begin{model}
\label{Chap2Model2}
Consider the class of non-linear SSMs with additive Gaussian noise
\begin{subequations}
 \label{eq:Chap2E23}
 \begin{align}
    {X}_{t+1} =&{f}_t({X}_{t})+ {V}_{t},	\label{eq:Chap2E23a}\\
    {Y}_t =&{g}_t({X}_{t})+ {W}_{t},\label{eq:Chap2E23b}
 \end{align}
 \end{subequations}
where ${V_t\in\mathbb{R}^n}$ and ${W_t\in\mathbb{R}^m}$ are mutually independent sequences from the Gaussian distribution, such that ${V_t\sim\mathcal{N}(v_t|0,Q_t)}$ and ${W_t\sim\mathcal{N}(w_t|0,R_t)}$.
\end{model}
Note that Model \ref{Chap2Model2} can also be represented as
\begin{subequations}
\begin{align}
\log[p&(X_{t+1}|X_t)]=c_1-\frac{1}{2}[X_{t+1}-f_t(X_t)]^TQ_t^{-1}\nonumber\\
&\times[X_{t+1}-f_t(X_t)],\label{eq:Chap2E25a}\\
\log[p&(Y_{t+1}|X_{t+1})]=c_2-\frac{1}{2}[Y_{t+1}-g_{t+1}(X_{t+1})]^TR_{t+1}^{-1}\nonumber\\
&\times[Y_{t+1}-g_{t+1}(X_{t+1})],\label{eq:Chap2E25b}
\end{align}
\end{subequations}
where ${c_1\in\mathbb{R}_+}$ and ${c_2\in\mathbb{R}_+}$ are normalizing constant and ${\mathbb{R}_+:=[0,\infty)}$.
\begin{result}
\label{result:Chap2R1}
The first and second order partial derivative of (\ref{eq:Chap2E25a}) is given by
\begin{subequations}
\label{eq:Chap2E26}
\begin{align}
\nabla_{X_t}\log[p(X_{t+1}|X_t)]=&[\nabla_{X_t}f^T_t(X_t)]Q_t^{-1}[X_{t+1}-f_t(X_t)],\label{eq:Chap2E26a}\\
\Delta_{X_t}^{X_t}\log[p(X_{t+1}|X_t)]=&-[\nabla_{X_t}f^T_t(X_t)]Q_t^{-1}[\nabla_{X_t}f_t(X_t)]\nonumber\\
&+[\Delta_{X_t}^{X_t}f^T_t(X_t)]\Lambda_{X_{t}}^{-1}\Psi_{X_{t}},\label{eq:Chap2E26b}
\end{align}
and the first with respect to ${X_{t+1}\in\mathcal{X}}$ and the second with respect to ${X_t\in\mathcal{X}}$ is given by
\begin{align}
\Delta_{X_t}^{X_{t+1}}\log[p(X_{t+1}|X_t)]=&[\nabla_{X_t}f^T_t(X_t)]Q_t^{-1},\label{eq:Chap2E26c}
\end{align}
\end{subequations}
where: ${\Lambda_{X_{t}}^{-1}=Q_t^{-1}I_{n^2\times n^2}}$; ${\Psi_{X_{t}}=[X_{t+1}-f_t(X_t)]I_{n^2\times n}}$; $I_{n^2\times n^2}$, and $I_{n^2\times n}$ are ${n^2\times n^2}$ and ${n^2\times n}$ identity matrix, respectively.  Also: $[\nabla_{X_t}f^T_t(X_t)]$ and $[\Delta_{X_t}^{X_t}f^T_t(X_t)]$ are
\begin{subequations}
\begin{align}
[\nabla_{X_t}f^T_t(X_t)]\triangleq&[\nabla_{X_t}f^{(1)}_t(X_t),\cdots, \nabla_{X_t}f^{(n)}_t(X_t)]_{n\times n},\label{eq:Chap2E28a}\\
[\Delta_{X_t}^{X_t}f^T_t(x_t)]\triangleq&[\Delta_{X_t}^{X_t}f^{(1)}_t(X_t),\cdots, \Delta_{X_t}^{X_t}f^{(n)}_t(X_t)]_{n\times n^2},\label{eq:Chap2E28b}
\end{align}
\end{subequations}
where $f_t(X_t)\triangleq[f^{(1)}_t(X_t),\cdots,f^{(n)}_t(X_t)]^T$ is a $n\times1$ vector valued function in (\ref{eq:Chap2E23a}).
\end{result}
\begin{result}
\label{result:Chap2R2}
The second order partial derivative of (\ref{eq:Chap2E25a}) and (\ref{eq:Chap2E25b}) is given by
\begin{subequations}
\label{eq:Chap2EF3}
\begin{align}
&\Delta_{X_{t+1}}^{X_{t+1}}\log[p(X_{t+1}|X_t)]=-Q_t^{-1}\label{eq:Chap2EF3a}\\
&\Delta_{X_{t+1}}^{X_{t+1}}\log[p(Y_{t+1}|X_{t+1})]=[\Delta_{X_{t+1}}^{X_{t+1}}g^T_{t+1}(X_{t+1})]\Lambda_{Y_{t+1}}^{-1}\Psi_{Y_{t+1}}\nonumber\\
&-[\nabla_{X_{t+1}}g^T_{t+1}(X_{t+1})]R_{t+1}^{-1}[\nabla_{X_{t+1}}g_{t+1}(X_{t+1})]\label{eq:Chap2EF3b}
\end{align}
\end{subequations}
where: ${\Lambda_{Y_{t+1}}^{-1}=R_{t+1}^{-1}I_{n^2\times n^2}}$; ${\Psi_{Y_{t+1}}=[Y_{t+1}-g_{t+1}(X_{t+1})]I_{n^2\times n}}$; ${I_{n^2\times n^2}}$, and ${I_{n^2\times n}}$ are ${n^2\times n^2}$ and ${n^2\times n}$ identity matrix. Also: ${[\nabla_{X_{t+1}}g_{t+1}(X_{t+1})]}$ and ${[\Delta_{X_{t+1}}^{X_{t+1}}g_{t+1}(X_{t+1})]}$ are
\begin{subequations}
\begin{align}
&[\nabla_{X_{t+1}}g^T_{t+1}(X_{t+1})]=\nonumber\\
&=[\nabla_{X_{t+1}}g^{(1)}_{t+1}(X_{t+1}),\dots,\nabla_{X_{t+1}}g^{(m)}_{t+1}(X_{t+1})]_{m\times m}\label{eq:Chap2EF5a};\\
&[\Delta_{X_{t+1}}^{X_{t+1}}g^T_{t+1}(X_{t+1})]\nonumber\\
&=[\Delta_{X_{t+1}}^{X_{t+1}}g^{(1)}_{t+1}(X_{t+1}),\dots, \Delta_{X_{t+1}}^{X_{t+1}}g^{(m)}_{t+1}(X_{t+1})]_{m\times m^2}\label{eq:Chap2EF5b};
\end{align}
\end{subequations}
where ${g_{t+1}(X_{t+1})\triangleq[g^{(1)}_{t+1}(X_{t+1}),\cdots,g^{(n)}_{t+1}(X_{t+1})]^T}$ is a ${m\times1}$ vector function in (\ref{eq:Chap2E23b}).
\end{result}
\begin{lemma}
\label{lemma:Chap2L8}
For a system given by Model \ref{Chap2Model2}, under Assumptions \ref{Chap2A1} through \ref{Chap2A3} the matrices (\ref{eq:Chap2EQE1a}) through (\ref{eq:Chap2EQE1d}) in Lemma \ref{lemma:Chap2L3} can be written as:
\begin{subequations}
\begin{align}
I_t^{11}=&\mathbb{E}_{p(X_{t}|Y_{1:t+1})}[\nabla_{X_t}f^T_t(X_t)]Q_t^{-1}[\nabla_{X_t}f_t(X_t)];\label{eq:Chap2E30a}\\
I_t^{12}=&\mathbb{E}_{p(X_t|Y_{1:t+1})}[-\nabla_{X_t}f^T_t(X_t)]Q_t^{-1};\label{eq:Chap2E30b}\\
I_t^{22,a}=&Q_t^{-1};\label{eq:Chap2E30c}\\
I_t^{22,b}=&\mathbb{E}_{\frac{p(Y_{1:t})}{p(Y_{1:t+1})}}\mathbb{E}_{p(X_{t+1}|Y_{1:t})}[\nabla_{X_{t+1}}g^T_{t+1}(X_{t+1})]R_{t+1}^{-1}\nonumber\\
&\times[\nabla_{X_{t+1}}g^T_{t+1}(X_{t+1})].\label{eq:Chap2E30d}
\end{align}
\end{subequations}
\end{lemma}
\begin{IEEEproof} (\ref{eq:Chap2E30a}):  Substituting (\ref{eq:Chap2E26b}) into (\ref{eq:Chap2EQE1a}) yields
\begin{subequations}
\label{eq:Chap2E29}
\begin{align}
&I_t^{11}=\mathbb{E}_{p(X_{t:t+1}|Y_{1:t+1})}[[\nabla_{X_t}f^T_t(X_t)]Q_t^{-1}[\nabla_{X_t}f_t(X_t)]\nonumber\\
&-[\Delta_{X_t}^{X_t}f^T_t(x_t)]\Lambda_{X_{t}}^{-1}\Psi_{X_{t}}],\label{eq:Chap2E29a}\\
&=\mathbb{E}_{p(X_{t}|Y_{1:t+1})}\mathbb{E}_{p(X_{t+1}|X_t, Y_{1:t+1})}[[\nabla_{X_t}f^T_t(X_t)]Q_t^{-1}\nonumber\\
&\times[\nabla_{X_t}f_t(X_t)]-[\Delta_{X_t}^{X_t}f^T_t(X_t)]\Lambda_{X_{t}}^{-1}\Psi_{X_{t}}],\label{eq:Chap2E29b}
\end{align}
\end{subequations}
where (\ref{eq:Chap2E29b}) is obtained by substituting the probability relation $p(x_{t:t+1}|y_{1:t+1})=p(x_{t+1}|x_t, y_{1:t+1})p(x_{t}|y_{1:t+1})$ into (\ref{eq:Chap2E29a}). Finally, by noting the following two conditions
\begin{align}
&\mathbb{E}_{p(X_{t+1}|X_t,Y_{1:t+1})}[\nabla_{X_t}f^T_t(X_t)]Q_t^{-1}[\nabla_{X_t}f_t(X_t)]\nonumber\\
&=[\nabla_{X_t}f^T_t(X_t)]Q_t^{-1}[\nabla_{X_t}f_t(X_t)],\label{eq:Chap2E29c}\\
&\mathbb{E}_{p(X_{t+1}|X_t,Y_{1:t+1})}[\Delta_{X_t}^{X_t}f^T_t(X_t)]\Lambda_{X_{t}}^{-1}\Psi_{X_{t}}\nonumber\\
&=[\Delta_{X_t}^{X_t}f^T_t(X_t)]\Lambda_{X_{t}}^{-1}\mathbb{E}_{p(X_{t+1}|X_t,Y_{1:t+1})}[\Psi_{x_{t}}]=0,\label{eq:Chap2E29d}
\end{align}
and substituting (\ref{eq:Chap2E29c}) and (\ref{eq:Chap2E29d}) into (\ref{eq:Chap2E29b}) yields (\ref{eq:Chap2E30a}).\\
\noindent
(\ref{eq:Chap2E30b}): Substituting (\ref{eq:Chap2E26c}) into (\ref{eq:Chap2EQE1b}) yields
\begin{align}
\label{eq:Chap2EAF2}
I_t^{12}=\mathbb{E}&_{p(X_{t:t+1}|Y_{1:t+1})}[-[\nabla_{X_t}f^T_t(X_t)]Q_t^{-1}].
\end{align}
Substituting the probability relation $p(x_{t:t+1}|y_{1:t+1})=p(x_{t+1}|x_t, y_{1:t+1})p(x_{t}|y_{1:t+1})$ into (\ref{eq:Chap2EAF2}), followed by taking independent terms out of the integral yields (\ref{eq:Chap2E30b}).\\
\noindent
(\ref{eq:Chap2E30c}): Substituting (\ref{eq:Chap2EF3a}) into (\ref{eq:Chap2EQE1c}) yields (\ref{eq:Chap2E30c}).\\
\noindent
(\ref{eq:Chap2E30d}): Using Bayes' rule, the expectation in (\ref{eq:Chap2EQE1d}) can be rewritten as
\begin{align}
I_t^{22,b}=\mathbb{E}_{\frac{p(X_{t+1}, Y_{1:t+1})}{p(Y_{1:t+1})}}[-\Delta_{X_{t+1}}^{X_{t+1}}\log~p(Y_{t+1}|X_{t+1})].\label{eq:Chap2EF9a}
\end{align}
Now using the probability condition ${p(x_{t+1},y_{1:t+1})=p(y_{t+1}|x_{t+1})p(x_{t+1}|y_{1:t})p(y_{1:t})}$, the expectation in (\ref{eq:Chap2EF9a}) can further be decomposed and written as
\begin{align}
I_t^{22,b}=&\mathbb{E}_{\frac{p(Y_{1:t})}{p(Y_{1:t+1})}}\mathbb{E}_{p(X_{t+1}|Y_{1:t})}\mathbb{E}_{p(Y_{t+1}|X_{t+1})}\nonumber\\
&\times[-\Delta_{X_{t+1}}^{X_{t+1}}\log~p(Y_{t+1}|X_{t+1})].\label{eq:Chap2EF9b}
\end{align}
Substituting (\ref{eq:Chap2EF3b}) into (\ref{eq:Chap2EF9b}) yields
\begin{align}
\label{eq:Chap2EF10}
&I_t^{22,b}=\mathbb{E}_{\frac{p(Y_{1:t})}{p(Y_{1:t+1})}}\mathbb{E}_{p(X_{t+1}|Y_{1:t})}\mathbb{E}_{p(Y_{t+1}|X_{t+1})}\nonumber\\
&\left[[-\Delta_{X_{t+1}}^{X_{t+1}}g^T_{t+1}(X_{t+1})]\Lambda_{Y_{t+1}}^{-1}\Psi_{Y_{t+1}}\right.\nonumber\\
&\left.+[\nabla_{X_{t+1}}g^T_{t+1}(X_{t+1})]R_{t+1}^{-1}[\nabla_{X_{t+1}}g_{t+1}(X_{t+1})]\right].
\end{align}
Noting the following two conditions
\begin{align}
&\mathbb{E}_{p(Y_{t+1}|X_{t+1})}[\nabla_{X_{t+1}}g^T_{t+1}(X_{t+1})]R_{t+1}^{-1}[\nabla_{X_{t+1}}g_{t+1}(X_{t+1})]\nonumber\\
&=[\nabla_{X_{t+1}}g^T_{t+1}(X_{t+1})]R_{t+1}^{-1}[\nabla_{X_{t+1}}g_{t+1}(X_{t+1})],\label{eq:Chap2EF11}\\
&\mathbb{E}_{p(Y_{t+1}|X_{t+1})}[\Delta_{X_{t+1}}^{X_{t+1}}g^T_{t+1}(X_{t+1})]\Lambda_{Y_{t+1}}^{-1}\Psi_{Y_{t+1}}]\nonumber\\
&=\Delta_{X_{t+1}}^{X_{t+1}}g^T_{t+1}(X_{t+1})\Lambda_{Y_{t+1}}^{-1}\mathbb{E}_{p(Y_{t+1}|X_{t+1})}[\Psi_{Y_{t+1}}]=0,\label{eq:Chap2EF12}
\end{align}
and substituting (\ref{eq:Chap2EF11}) and (\ref{eq:Chap2EF12}) into (\ref{eq:Chap2EF10}) yields (\ref{eq:Chap2E30d}), which completes the proof.
\end{IEEEproof}
Using the results of Lemma \ref{lemma:Chap2L8}, an SMC approximation of the PCRLB for Model \ref{Chap2Model2} can be subsequently computed, as discussed in the next lemma.

\begin{lemma}
\label{lemma:Chap2L9}
Let a stochastic non-linear system with additive Gaussian state and sensor noise be represented by Model \ref{Chap2Model2}, such that it satisfies Assumption \ref{Chap2A1} through \ref{Chap2A3}. Let ${\{Y_{1:t}=y^j_{1:t}\}_{j=1}^M}$ be ${M\in\mathbb{N}}$ i.i.d.~measurement sequences generated from Model \ref{Chap2Model2}, then (\ref{eq:Chap2E5b}) through (\ref{eq:Chap2E5e}) in Lemma \ref{lemma:Chap2L2} can be recursively approximated as follows:
\begin{subequations}
\begin{align}
\tilde{D}_t^{11}=&\frac{1}{MN}\sum_{j=1}^M\sum_{i=1}^N[\nabla_{X_t}f^T_t(X^{i,j}_{t|t+1})]Q_t^{-1}[\nabla_{X_t}f_t(X^{i,j}_{t|t+1})];\label{eq:Chap2EQ38a}\\
\tilde{D}_t^{12}=&\frac{1}{MN}\sum_{j=1}^M\sum_{i=1}^N-[\nabla_{X_t}f^T_t(X^{i,j}_{t|t+1})]Q_t^{-1};\label{eq:Chap2EQ38b}\\
\tilde{D}_t^{22}=&Q_t^{-1}+\frac{1}{MN}\sum_{j=1}^M\sum_{i=1}^N[\nabla_{X_{t+1}}g^T_{t+1}(X^{i,j}_{t+1|t})]R_{t+1}^{-1}\nonumber\\
&\times[\nabla_{X_{t+1}}g^T_{t+1}(X^{i,j}_{t+1|t})]; \label{eq:Chap2EQ38c}
\end{align}
\end{subequations}
and ${\{X^{i,j}_{t|t+1}\}_{i=1}^{N}\sim p(x_{t}|y^j_{1:t+1})}$ and ${\{X^{i,j}_{t+1|t}\}_{i=1}^{N}\sim p(x_{t+1}|y^j_{1:t})}$ are sets of $N$ resampled particles from Lemma \ref{lemma:Chap2L6} and {Algorithm \ref{algorithm:Chap2A0}}, respectively, for all ${\{Y_{1:t+1}=y^j_{1:t+1}\}_{j=1}^M}$. 
\end{lemma}
\begin{IEEEproof}
For ${\{Y_{1:t}=y^j_{1:t}\}}$, the SMC approximation in (\ref{eq:Chap2EL5}) can be written as
\begin{align}
\label{eq:Chap2EQ39}
\tilde{p}(dx_{t}|y^j_{1:t+1})=\frac{1}{N}\sum_{i=1}^N\delta_{X^{i,j}_{t|t+1}}(dx_{t}),
\end{align}
where ${X^{i,j}_{t|t+1}\sim p(x_{t}|y^j_{1:t+1})}$. Substituting (\ref{eq:Chap2EQ39}) into (\ref{eq:Chap2E30a}) and (\ref{eq:Chap2E30b}) yields
\begin{subequations}
\label{eq:Chap2EQ40}
\begin{align}
\tilde{I}_t^{11}=&\frac{1}{N}\sum_{i=1}^N[\nabla_{X_t}f^T_t(X^{i,j}_{t|t+1})]Q_t^{-1}[\nabla_{X_t}f_t(X^{i,j}_{t|t+1})],\label{eq:Chap2EQ40a}\\
\tilde{I}_t^{12}=&-\frac{1}{N}\sum_{i=1}^N[\nabla_{X_t}f^T_t(X^{i,j}_{t|t+1})]Q_t^{-1},\label{eq:Chap2EQ40b}
\end{align}
\end{subequations} 
where $\tilde{I}_t$ is an SMC approximations of ${I}_t$. Substituting  (\ref{eq:Chap2EQ40}) and (\ref{eq:Chap2EQ9}) into (\ref{eq:Chap2E6e}) and (\ref{eq:Chap2E6f}) yields (\ref{eq:Chap2EQ38a}) and (\ref{eq:Chap2EQ38b}), respectively. Computing an SMC approximation of $D_t^{22}$ in (\ref{eq:Chap2E6g}) for Model \ref{Chap2Model2} requires a slightly different approach. Substituting (\ref{eq:Chap2E30c}) and (\ref{eq:Chap2E30d}) into (\ref{eq:Chap2E6g}) yields
\begin{subequations}
\label{eq:Chap2EQ41}
\begin{align}
D_t^{22}=&\mathbb{E}_{p(Y_{1:t+1})}[Q_t^{-1}+\mathbb{E}_{\frac{p(Y_{1:t})}{p(Y_{1:t+1})}}\mathbb{E}_{p(X_{t+1}|Y_{1:t})}\nonumber\\
&\times[\nabla_{X_{t+1}}g^T_{t+1}(X_{t+1})]R_{t+1}^{-1}[\nabla_{X_{t+1}}g^T_{t+1}(X_{t+1})]],\label{eq:Chap2EQ41a}\\
=&Q_t^{-1}+\mathbb{E}_{p(Y_{1:t})}\mathbb{E}_{p(X_{t+1}|Y_{1:t})}[\nabla_{X_{t+1}}g^T_{t+1}(X_{t+1})]R_{t+1}^{-1}\nonumber\\
&\times[\nabla_{X_{t+1}}g^T_{t+1}(X_{t+1})],\label{eq:Chap2EQ41b}
\end{align}
\end{subequations}
where $Q_t^{-1}$ is independent of the measurement sequence. Also, $\mathbb{E}_{p(Y_{1:t+1})}\mathbb{E}_{\frac{p(Y_{1:t})}{p(Y_{1:t+1})}}[\cdot]=\mathbb{E}_{p(Y_{1:t})}[\cdot]$. For ${\{Y_{1:t}=y^j_{1:t}\}}$, random samples $\{X^{i,j}_{t+1|t}\}_{i=1}^N\sim p(x_{t+1}|y^j_{1:t})$ from {Algorithm \ref{algorithm:Chap2A0}} delivers an SMC approximation of $p(dx_{t+1}|y^j_{1:t})$ given as  
\begin{align}
\label{eq:Chap2EQ42}
\tilde p(dx_{t+1}|y^j_{1:t})=\frac{1}{N}\sum_{i=1}^N\delta_{X^{i,j}_{t+1|t}}(dx_{t+1})
\end{align}
where $\tilde p(dx_{t+1}|y^j_{1:t})$ is an SMC approximation of $p(dx_{t+1}|y^j_{1:t})$. Substituting (\ref{eq:Chap2EQ42}) and (\ref{eq:Chap2EQ9}) into (\ref{eq:Chap2EQ41b}) yields (\ref{eq:Chap2EQ38c}), which completes the proof.
\end{IEEEproof}
\begin{result}
\label{result:Chap2R1E}
An SMC approximation of the PFIM for Model \ref{Chap2Model2} is obtained by substituting (\ref{eq:Chap2EQ38a}) through (\ref{eq:Chap2EQ38c}) in Lemma \ref{lemma:Chap2L9} into (\ref{eq:Chap2E4}) in Lemma \ref{lemma:Chap2L2}, such that
\begin{align}
\label{eq:Chap2E22}
\tilde{J}_{t+1}=\tilde{D}_t^{22}-[\tilde{D}_t^{12}]^T(\tilde{J}_t+\tilde{D}_t^{11})^{-1}\tilde{D}_t^{12},
\end{align}
where $\tilde{J}_{t+1}$ is an SMC approximation of ${J}_{t+1}$. Applying matrix inversion lemma \cite{RA1985} in (\ref{eq:Chap2E22}) gives an SMC approximation of the PCRLB, such that
\begin{align}
\label{eq:Chap2EQ11}
&\tilde{J}_{t+1}^{-1}=[\tilde{D}_t^{22}]^{-1}-[\tilde{D}_t^{22}]^{-1}[\tilde{D}_t^{12}]^T\nonumber\\
&\times\left[\tilde{D}_t^{12}[\tilde{D}_t^{22}]^{-1}[\tilde{D}_t^{12}]^T-(\tilde{J}_t+\tilde{D}_t^{11})\right]^{-1}\tilde{D}_t^{12}[\tilde{D}_t^{22}]^{-1},
\end{align}
where $\tilde{J}_{t+1}^{-1}$ is an SMC approximation of ${J}_{t+1}^{-1}$ in (\ref{eq:Chap2E2}).
\end{result}
\section{Final Algorithm}
\label{sec:Chap2S5}
Algorithms \ref{algorithm:Chap2A1} and \ref{algorithm:Chap2A2} give the procedure for computing an SMC approximation of the PCRLB for Models \ref{Chap2Model1} and \ref{Chap2Model2}, respectively.
\begin{algorithm}[h]
  \caption{SMC based PCRLB for Model \ref{Chap2Model1}}
  \label{algorithm:Chap2A1}
  \begin{algorithmic}[1]
	\LCOMMENT ~ \textbf{Input:} Given Model \ref{Chap2Model1}, satisfying Assumptions \ref{Chap2A1} through \ref{Chap2A3}, assume a prior pdf on $X_0$, such that ${X_0\sim p(x_0)}$. Also, select algorithm parameters- $T$, $N$ and $M$.
	\LCOMMENT ~ \textbf{Output:} SMC approximation of the PCRLB for Model \ref{Chap2Model1}.
    \STATE Generate and store $M$ i.i.d.~sequences ${{\{Y^j_{1:T}\}_{j=1}^M}\sim p(y_{1:T})}$ of length T, by simulating Model \ref{Chap2Model1}, $M$ times starting at $M$ i.i.d.~initial states ${\{X^i_{0|-1}\}_{j=1}^M\sim p(x_0)}$.
    \FOR{$j=1~\text{to}~M$}    
    \FOR{$t=1~\text{to}~T$}
    \STATE Store resampled particles ${\{{X}^{i,j}_{t|t}\}_{i=1}^{N}\sim p(x_t|y^j_{1:t})}$  using {Algorithm \ref{algorithm:Chap2A0}}.
    \STATE Store resampled particles ${\{{X}^{i,j}_{t-1:t|t}\}_{i=1}^{N}\sim p(x_{t-1:t}|y^j_{1:t})}$ using Lemma \ref{lemma:Chap2L5}.
    \ENDFOR
    \ENDFOR
    \STATE Compute PFIM $J_0$ at ${t=0}$ based on the initial target state pdf ${X_0\sim p(x_0)}$. If $X_0\sim \mathcal{N}(x_0|C_{x_{0}},P_{0|0})$ then from Lemma \ref{lemma:Chap2L2}, ${J_0=P^{-1}_{0|0}}$.
    \FOR{$t=0~\text{to}~T-1$}
    \STATE Compute an SMC estimate (\ref{eq:Chap2EE38a}) through (\ref{eq:Chap2EE38c}) in Lemma \ref{lemma:Chap2L7}.
    \STATE Compute PCRLB $\tilde{J}^{-1}_{t+1}$ by substituting (\ref{eq:Chap2EE38a}) through (\ref{eq:Chap2EE38c}) into (\ref{eq:Chap2EQ11}). 
    \ENDFOR    
\end{algorithmic}
\end{algorithm}
\begin{algorithm}[h]
  \caption{SMC based PCRLB for Model \ref{Chap2Model2}}
  \label{algorithm:Chap2A2}
  \begin{algorithmic}[1]
	\LCOMMENT ~ \textbf{Input:} Given Model \ref{Chap2Model2}, satisfying Assumptions \ref{Chap2A1} through \ref{Chap2A3}, assume a prior on $X_0$, such that ${X_0\sim p(x_0)}$. Also, select algorithm parameters- $T$, $N$ and $M$.
	\LCOMMENT ~ \textbf{Output:} SMC approximation of the PCRLB for Model \ref{Chap2Model2}.
    \STATE Generate and store $M$ i.i.d.~sequences ${{\{Y^j_{1:T}\}_{j=1}^M}\sim p(y_{1:T})}$ of length T, by simulating Model \ref{Chap2Model2}, $M$ times starting at $M$ i.i.d.~initial states ${\{X^i_{0|-1}\}_{j=1}^M\sim p(x_0)}$.
    \FOR{$j=1~\text{to}~M$}    
    \FOR{$t=1~\text{to}~T$}
    \STATE Store predicted particles ${\{{X}^{i,j}_{t|t-1}\}_{i=1}^{N}\sim p(x_t|y^j_{1:t-1})}$  using {Algorithm \ref{algorithm:Chap2A0}}.
    \STATE Store resampled particles ${\{{X}^{i,j}_{t|t}\}_{i=1}^{N}\sim p(x_t|y^j_{1:t})}$  using {Algorithm \ref{algorithm:Chap2A0}}.
    \STATE Store resampled particles ${\{{X}^{i,j}_{t-1|t}\}_{i=1}^{N}\sim p(x_{t-1:t}|y^j_{1:t})}$ using Lemma \ref{lemma:Chap2L6}.
    \ENDFOR
    \ENDFOR
    \STATE Compute PFIM $J_0$ at ${t=0}$ based on the initial target state pdf ${X_0\sim p(x_0)}$. If $X_0\sim \mathcal{N}(x_0|C_{x_{0}},P_{0|0})$ then from Lemma \ref{lemma:Chap2L2}, ${J_0=P^{-1}_{0|0}}$.
    \FOR{$t=0~\text{to}~T-1$}
    \STATE Compute an SMC estimate (\ref{eq:Chap2EQ38a}) through (\ref{eq:Chap2EQ38c}) in Lemma \ref{lemma:Chap2L9}.
    \STATE Compute PCRLB $\tilde{J}^{-1}_{t+1}$ by substituting (\ref{eq:Chap2EQ38a}) through (\ref{eq:Chap2EQ38c}) into  using (\ref{eq:Chap2EQ11}). 
    \ENDFOR    
\end{algorithmic}
\end{algorithm}
\begin{rmk}
\label{remark:Chap2R8}
In practice, an ensemble of $M$ measurement sequences ${\{Y_{1:T}=y^j_{1:T}\}_{j=1}^M}$ required by Algorithms \ref{algorithm:Chap2A1} and \ref{algorithm:Chap2A2} are obtained from historical process data; however, in simulations, it can be generated by simulating Models \ref{Chap2Model1} and \ref{Chap2Model2}, $M$ times starting at i.i.d.~initial states drawn from ${X_0\sim p(x_0)}$. Note that this procedure also requires simulation of the true states; however, true states are not used in Algorithms \ref{algorithm:Chap2A1} and \ref{algorithm:Chap2A2}.  
\end{rmk}
For illustrative purposes, to assess the numerical reliability of Algorithms \ref{algorithm:Chap2A1} and \ref{algorithm:Chap2A2}, a quality measure is defined as follows
\begin{align}
\label{eq:Chap2Extra1}
\Lambda_J=\frac{1}{T}\sum_{t=1}^T[{J}_t^{-1}-\tilde{J}_t^{-1}]\circ[{J}_t^{-1}-\tilde{J}_t^{-1}],
\end{align}
where $\Lambda_J$ is the average sum of square of errors in approximating the PCRLB and $\circ$ is the Hadamard product. $\Lambda_J$ is a $n\times n$ matrix, with diagonal element $\Lambda_J(j,j)$ as the average sum of square of errors accumulated in approximating the PCRLB for state $j$, where $1\leq j\leq n$.
\section{Convergence}
\label{sec:Chap2S6}
Computing the PCRLB in  Lemma \ref{lemma:Chap2L1} involves solving the complex, multi-dimensional integrals; however, as stated earlier, for Models \ref{Chap2Model1} and \ref{Chap2Model2} the PCRLB cannot be solved in closed form. Algorithms \ref{algorithm:Chap2A1} and \ref{algorithm:Chap2A2} gives a $N$ particle and $M$ simulation based SMC approximation of the PCRLB for Models \ref{Chap2Model1} and \ref{Chap2Model2}, respectively. It is therefore natural to question the convergence properties of the proposed numerical method. In this regard, results such as Theorem \ref{theorem:Chap2T2} and Remark \ref{remark:Chap2R6} are important as it ensures that the proposed numerical solution does not result in accumulation of errors. It is emphasized that although Theorem \ref{theorem:Chap2T2} and Remark \ref{remark:Chap2R6} not necessarily imply convergence of the SMC based PCRLB and MSE to its theoretical values, nevertheless, it provides a strong theoretical basis for the numerous approximations used in Algorithms \ref{algorithm:Chap2A1} and \ref{algorithm:Chap2A2}. 

From an application perspective, it is instructive to highlight that the numerical quality of the SMC based PCRLB approximation in Algorithms \ref{algorithm:Chap2A1} and \ref{algorithm:Chap2A2} can be made accurate by simply increasing the number of particles ($N$) and the MC simulations ($M$). The choice of $N$ and $M$ are user defined, which can be selected based on the required numerical accuracy, and available computing speed. It is important to emphasize that due to the multiple approximations involved in deriving a tractable solution, for practical purposes, with a finite $N$ and $M$, the condition ${P}_{t|t}-\tilde{J}^{-1}_t\succcurlyeq 0$ is not guaranteed to hold for all $t\in\mathbb{N}$. 

The quality of the SMC based PCRLB solution is validated next via simulation.
\section{Simulation examples}
\label{sec:Chap2S7}
In this section, two simulation examples are presented to demonstrate the utility and performance of the proposed SMC based PCRLB solution. The first example is a ballistic target tracking problem at re-entry phase. The aim of this study is three fold: first to demonstrate the performance and utility of the proposed method on a practical problem;  second, to demonstrate the quality of the bound approximation for a range of target state and sensor noise variances; and third, to study the sensitivity of the involved SMC approximations to the number of particles used.

The performance of the SMC based PCRLB solution on a second example involving a uni-variate, non-stationary growth model, which is a standard non-linear, and bimodal benchmark model is then illustrated. This example is profiled to demonstrate the accuracy of the SMC based PCRLB solution for highly non-linear SSMs with non-Gaussian noise.
\subsection{Example 1: Ballistic target tracking at re-entry}
\label{sec:Chap2S7.1}
In Section \ref{sec:Chap2S4.3}, an SMC based method for approximating the PCRLB was presented for non-linear SSMs with additive Gaussian state and sensor noise (See Algorithm \ref{algorithm:Chap2A2}). In this section, the quality of Algorithm \ref{algorithm:Chap2A2} is validated on a practical problem of ballistic target tracking at re-entry phase. This particular problem has attracted a lot of attention from researchers for both theoretical and practical reasons. See \cite{RP2001} and the references cited therein for a detailed survey on the ballistic target tracking. 
\subsubsection{Model setup}
Consider a target launched along a ballistic flight whose kinematics are described in a 2D Cartesian coordinate system. This particular description of the kinematics assumes that the only forces acting on the target at any given time are the forces due to gravity and drag. All other forces such as: centrifugal acceleration, Coriolis acceleration, wind, lift force and spinning motion are assumed to have a small effect on the target trajectory. With the position and the velocity of the target at time ${t\in\mathbb{N}}$ described  in 2D Cartesian coordinate system as $(\text{X}_t, \text{H}_t)$ and $(\dot{\text{X}}_t, \dot{\text{H}}_t)$, respectively, its motion in the re-entry phase can be described by the following discrete-time non-linear SSM \cite{ARB2002}
\begin{align}
\label{eq:Chap2E33}
{X}_{t+1}=A{X}_t+GF_t({X}_t)+G.\left[
\begin{array}{c}
0 \\
-g
\end{array}
\right]+V_t,
\end{align}
where the states ${X}_t\triangleq[\text{X}_t\quad \dot{\text{X}}_t\quad \text{H}_t\quad \dot{\text{H}}_t]^T$. Also, the matrices $A$ and $G$  are as follows
\begin{align}
A\triangleq\left[
  \begin{array}{cccc}
    1 & \Delta T & 0 & 0 \\
    0 & 1 & 0 & 0 \\
    0 & 0 & 1 & \Delta T \\
    0 & 0 & 0 & 1 \\
  \end{array}
\right],
G\triangleq
\left[
  \begin{array}{cc}
    \displaystyle {\frac{\Delta T^2}{2}} & 0 \\
      \displaystyle \Delta T & 0 \\
       0 &\displaystyle \frac{\Delta T^2}{2} \\
       0 &\displaystyle \Delta T \\
  \end{array}
\right],
\end{align}
where $\Delta T$ is the time interval between two consecutive radar measurements.

In (\ref{eq:Chap2E33}) $F_t(X_t)$ models the drag force, which acts in a direction opposite to the target velocity. In terms of the states, $F_t(X_t)$ can be modelled as
\begin{align}
\label{eq:Chap2E34}
\displaystyle{ F_t({X}_t)=-\frac{g\rho(\text{H}_t)}{2\beta}\sqrt{\dot{\text{X}}_t^2+\dot{\text{H}}_t^2}}
\left[
\begin{array}{c}
\dot{\text{X}}_t \\
\dot{\text{H}}_t
\end{array}
\right],
\end{align}
where: $g$ is the acceleration due to gravity; $\beta$ is the ballistic coefficient whose value depends on the shape, mass and the cross sectional area of the target \cite{R2004}; and $\rho(\text{H}_t)$ is the density of the air, defined as an exponentially decaying function of $\text{H}_t$, such that 
\begin{align}
\rho(\text{H}_t)=\alpha_1e^{(-\alpha_2\text{H}_t)}
\end{align}
where: $\alpha_1=1.227$ kg$\cdot \text{m}^{-3}$, $\alpha_2=1.09310\times 10^{-4}\text{m}^{-1}$ for $\text{H}_t<9144\text{m}$; and $\alpha_1=1.754$  kg$\cdot\text{m}^{-3}$, $\alpha_2=1.4910\times 10^{-4}\text{m}^{-1}$ for $\text{H}_t\geq9144 \text{m}$. Note that the drag force, $F_t(X_t)$ is the only non-linear term in the state equation. In (\ref{eq:Chap2E33}) the state noise ${V_t\in\mathbb{R}^4}$ is a i.i.d.~sequence of multi-variate Gaussian random vector represented as  ${V_t\sim \mathcal{N}(v_t|0,Q_t)}$, with zero mean and covariance matrix $Q_t$ given as
\begin{align}
Q_t=\gamma I_{2\times 2}\otimes \Theta,
\quad
\Theta=
\left[
\begin{array}{cc}
\displaystyle \frac{\Delta T^3}{3}&\displaystyle \frac{\Delta T^2}{2} \\\displaystyle
\displaystyle \frac{\Delta T^2}{2}&\displaystyle\Delta T
\end{array}
\right],\quad
\end{align}
where: $\gamma\in \mathbb{R}_+$; $I_{2\times 2}$ is a $2\times2$ identity matrix; and $\otimes$ is the Kronecker product. The intensity of the state noise, determined by $\gamma$, accounts for all the forces neglected in (\ref{eq:Chap2E33}), including any deviations arising due to system-model mismatch.  The target measurements are collected by a conventional radar (e.g., dish radar) assumed to be stationed at the origin. The sensor readings are measured in the natural sensor coordinate system, which include range ($R_t$) and elevation ($E_t$) of the target. The radar readings $Y_t=[R_t\quad E_t]^T$ are related to the states $X_t$ through a non-linear observation model given below.
\begin{align}
\label{eq:Chap2E35}
{Y}_{t}=\left[
\begin{array}{c}
\displaystyle{{\sqrt{\text{X}_t^2+\text{H}_t^2}}} \\
\displaystyle{\arctan\left(\frac{\text{H}_t}{\text{X}_t}\right)}
\end{array}
\right]+W_t.
\end{align}
In (\ref{eq:Chap2E35}) $W_t\in\mathbb{R}^2$ is an i.i.d. sequence of multi-variate Gaussian random vector represented as $W_t\sim \mathcal{N}(w_t|0,R_t)$, with zero mean and non-singular covariance matrix $R_t$ given as
\begin{align}
R_t=\left[
\begin{array}{cc}
\sigma_r^2& 0 \\
0&\sigma_e^2
\end{array}
\right],
\end{align}
where ${\sigma_r\in\mathbb{R}_+}$  and ${\sigma_e\in\mathbb{R}_+}$ are the standard deviation associated with range and elevation measurements. In (\ref{eq:Chap2E35}), it is assumed that the true target elevation angle lies between $0$ and $\pi/2$ radians; otherwise, it suffices to add $\pi$ radians to the $\arctan$ term in (\ref{eq:Chap2E35}).
\begin{rmk}
\label{remark:Chap2R9}
To avoid use of a non-linear sensor model, some authors \cite{ARB2002,Lei2011} considered transforming the radar measurements in (\ref{eq:Chap2E35}) into the Cartesian coordinate system, wherein the sensor dynamics manifest themselves into a linear model. Even though this strategy eliminates the need to handle non-linearity in sensor measurements, tracking in Cartesian coordinates couples the sensor noise across two coordinate systems and makes the noise non-Gaussian and state dependent \cite{RP2001c}. Since the proposed method can deal with strong state and sensor non-linearities, the radar readings are monitored in natural sensor coordinates alone.
\end{rmk}
\subsubsection{Simulation setup}
For simulation, the model parameters are selected as given in Table \ref{tab:Chap2Tab1}. The aim of this study is to evaluate the quality of the SMC based PCRLB solution for a range of target state and sensor noise variances. This allows full investigation of the quality of the SMC based approximation for a range of noise characteristics. The cases considered here are given in Table \ref{tab:Chap2Tab2}. From Assumption \ref{Chap2A1}, $\beta$ is assumed to be fixed and known {a priori}. 
\begin{table}[h]
\caption{Parameter values used in Example 1.}
\centering
\begin{tabular}{lll}
\hline
Process variables &Symbol&values\\
\hline
accel. due to gravity&$g$&$9.8~\text{m/s}^2$\\
ballistic coefficient &$\beta$&$40000~\text{kg.m}^{-1}\cdot\text{s}^{-2}$\\
radar sampling time&$\Delta T$&$2~\text{s}$\\
total tracking time&$T$&$120~\text{s}$\\
state noise&$V_t$&$V_t\sim \mathcal{N}(v_t|0,Q_t)$\\
sensor noise&$W_t$&$W_t\sim \mathcal{N}(w_t|0,R_t)$\\
noise parameters&$\gamma,~\sigma_r,~\sigma_e$&see Table \ref{tab:Chap2Tab2}\\
initial states&$X^\star_0$&$\left[
  \begin{array}{l}
  232~\text{km} \\
2.290\cos{(190^0)}~\text{km/s}\\
88~\text{km}\\
2.290\sin(190^o)~\text{km/s}\\
  \end{array}
\right]$\\
probability of detection&$\pr_d$&1\\
probability of false alarm&$\pr_f$&0\\
\hline
\end{tabular}
\label{tab:Chap2Tab1}
\end{table}
\begin{table}[h]
\caption{Cases considered for Example 1.}
\centering
\begin{tabular}{cccc}
\hline
Case &$\gamma$&$\sigma_r$&$\sigma_\epsilon$\\
\hline
1&$1.0$&$100\text{m}$&$0.017\text{rad}$\\
2&$5.0$&$100\text{m}$&$0.017\text{rad}$\\
3&$1.0$&$500\text{m}$&$0.085\text{rad}$\\
4&$5.0$&$500\text{m}$&$0.085\text{rad}$\\
\hline
\end{tabular}
\label{tab:Chap2Tab2}
\end{table}
\begin{figure*}[t]
   \centering
   \includegraphics[scale=0.6]{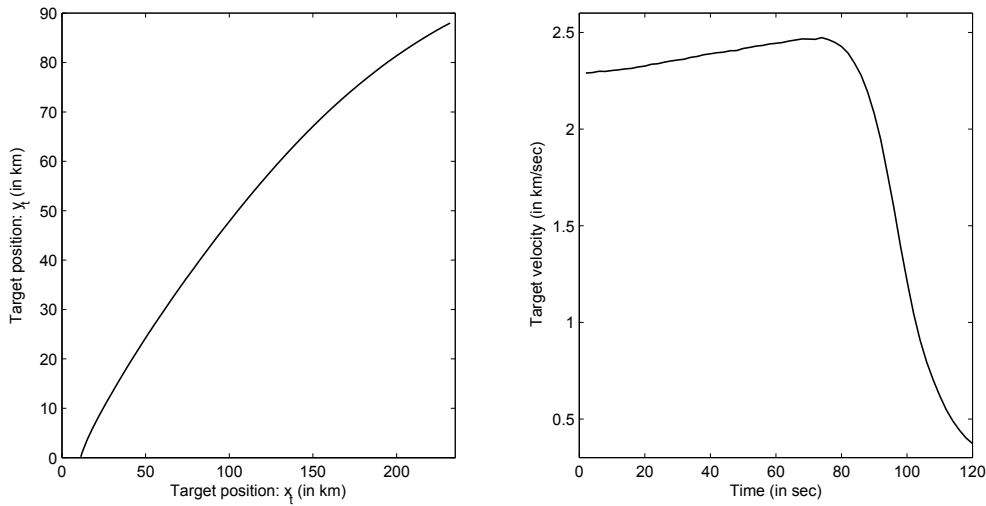}
   \caption{{Sample trajectory showing position and velocity of the target at re-entry phase.}}
   \label{fig:Chap2F0}
\end{figure*}
\begin{table}[!h]
\caption{Variable values used in Example 1.}
\centering
\begin{tabular}{lll}
\hline
Process variables &Symbol&values\\
\hline
state noise&$V_t$&$V_t\sim \mathcal{N}(v_t|0,Q_t)$\\
sensor noise&$W_t$&$W_t\sim \mathcal{N}(w_t|0,R_t)$\\
noise parameters&$\gamma,~\sigma_r,~\sigma_e$&see Table \ref{tab:Chap2Tab2}\\
initial states&$X_0$&$X_0\sim \mathcal{N}(x_0|C_{x_0},P_{0|0})$\\
&$C_{x_0}$&$\left[
  \begin{array}{l}
  232~\text{km} \\
2.290\cos{(190^0)}~\text{km/s}\\
88~\text{km}\\
2.290\sin(190^o)~\text{km/s}\\
  \end{array}
\right]$\\
&$[P_{0|0}]^{1/2}$&$\left[
  \begin{array}{llll}
    1\text{km} & 0 & 0 & 0 \\
    0 &20\text{m/s}  & 0 & 0 \\
    0 & 0 &1\text{km}& 0  \\
    0 & 0 & 0 & 20\text{m/s} \\
  \end{array}
\right]$\\
Number of particles&N&1000\\
MC simulations&M&200\\
\hline
\end{tabular}
\label{tab:Chap2Tab3}
\end{table}

Figure \ref{fig:Chap2F0} shows a sample trajectory of the target in the ${X-H}$ plane along with its velocity map as a function of time, generated using Case 1 (see Table \ref{tab:Chap2Tab2}).
\subsubsection{Results}
The kinematics of the ballistic target consist of nonlinear state and sensor models with additive Gaussian noise, for which the PCRLB can be approximated using Algorithm \ref{algorithm:Chap2A2}. First, the state and sensor models in (\ref{eq:Chap2E33}) and (\ref{eq:Chap2E35}), respectively, are defined as
\begin{subequations}
\begin{align}
f_t(X_t)&=A{X}_t+GF_t({X}_t)+G.\left[
\begin{array}{c}
0 \\
-g
\end{array}
\right]\label{eq:Chap2E36a},\\
g_{t+1}(X_{t+1})&=\left[
\begin{array}{c}
\displaystyle{{\sqrt{\text{X}_{t+1}^2+\text{H}_{t+1}^2}}} \\
\displaystyle{\arctan\left(\frac{\text{H}_{t+1}}{\text{X}_{t+1}}\right)}
\end{array}
\right].\label{eq:Chap2E36b}
\end{align}
\end{subequations}
To compute the required gradients $\nabla_{X_t}f_t(X_t)$ and $\nabla_{X_{t+1}}g_{t+1}(X_{t+1})$, differentiating (\ref{eq:Chap2E33}) with respect to $X_t$, and  (\ref{eq:Chap2E35}) with respect to $X_{t+1}$, yields
\begin{subequations}
\begin{align}
\nabla_{X_t}f_t(X_t)&=A+GM_t({X}_t),\label{eq:Chap2E37a}\\
\nabla_{X_{t+1}}g_{t+1}(X_{t+1})&=N_{t+1}({X}_{t+1})\label{eq:Chap2E37b},
\end{align}
\end{subequations}
where: $M_t({X}_t)$ and $N_{t+1}({X}_{t+1})$ in (\ref{eq:Chap2E37a}) and (\ref{eq:Chap2E37b}), respectively, are $2\times4$ matrices, whose entries are:
\begin{subequations}
\begin{align}
M_t(X_t)[1,1]&=0,\\
M_t(X_t)[2,1]&=0,\\
M_t(X_t)[1,2]&=-\frac{g}{2\beta}\rho(\text{H}_t)\left[\frac{2\dot{\text{X}}_t^2+\dot{\text{H}}_t^2}{\sqrt{\dot{\text{X}}_t^2+\dot{\text{H}}_t^2}}\right],\\
M_t(X_t)[2,2]&=-\frac{g}{2\beta}\rho(\text{H}_t)\left[\frac{\dot{\text{X}}_t\dot{\text{H}}_t}{\sqrt{\dot{\text{X}}_t^2+\dot{\text{H}}_t^2}}\right],\\
M_t(X_t)[1,3]&=\frac{g\alpha_2}{2\beta}\rho(\text{H}_t)\left[{\sqrt{\dot{\text{X}}_t^2+\dot{\text{H}}_t^2}}\right]\dot{\text{X}}_t,\\
M_t(X_t)[2,3]&=\frac{g\alpha_2}{2\beta}\rho(\text{H}_t)\left[{\sqrt{\dot{\text{X}}_t^2+\dot{\text{H}}_t^2}}\right]\dot{\text{H}}_t,
\end{align}
\begin{align}
M_t(X_t)[1,4]&=M_t(X_t)[2,2],\\
M_t(X_t)[2,4]&=-\frac{g}{2\beta}\rho(\text{H}_t)\left[\frac{\dot{\text{X}}_t^2+2\dot{\text{H}}_t^2}{\sqrt{\dot{\text{X}}_t^2+\dot{\text{H}}_t^2}}\right];
\end{align}
\end{subequations}
and:
\begin{subequations}
\begin{align}
N_{t+1}(X_{t+1})[1,1]&=\frac{\text{X}_{t+1}}{\sqrt{\dot{\text{X}}_{t+1}^2+\dot{\text{H}}_{t+1}^2}},\\
N_{t+1}(X_{t+1})[2,1]&=\frac{\text{H}_{t+1}}{{\dot{\text{X}}_{t+1}^2+\dot{\text{H}}_{t+1}^2}},\\
N_{t+1}(X_{t+1})[1,2]&=0,\\
N_{t+1}(X_{t+1})[2,2]&=0,\\
N_{t+1}(X_{t+1})[1,3]&=\frac{\text{H}_{t+1}}{\sqrt{\dot{\text{X}}_{t+1}^2+\dot{\text{H}}_{t+1}^2}},\\
N_{t+1}(X_{t+1})[2,3]&=\frac{\text{X}_{t+1}}{{\dot{\text{X}}_{t+1}^2+\dot{\text{H}}_{t+1}^2}},\\
N_{t+1}(X_{t+1})[1,4]&=0,\\
N_{t+1}(X_{t+1})[2,4]&=0.
\end{align}
\end{subequations}
To evaluate the numerical quality of Algorithm 1, we compare the SMC based PCRLB solution against the theoretical values. The theoretical bound is computed using an ensemble of the true state trajectories, simulated using (\ref{eq:Chap2E33}) (see \cite{ARB2002,R2004} for further details). Here we compare the square root of the diagonal elements of the theoretical PCRLB matrix $J^{-1}_t$ and its approximation $\tilde{J}^{-1}_t$ for all $t\in[0,T]$. The results are summarized next for the cases given in Table \ref{tab:Chap2Tab2}. For fair comparison of all the cases, the parameters required by Algorithm \ref{algorithm:Chap2A2} are specified as given in Table \ref{tab:Chap2Tab3}.

\begin{figure*}[!htbp]
    \label{fig:subfigures}
    \begin{center}
        \subfigure[Square root of the theoretical (solid line with marker) and approximate PCRLB (solid line) for all the target states under Case 1]{%
            \label{fig:Chap2F1}
            \includegraphics[scale=0.40]{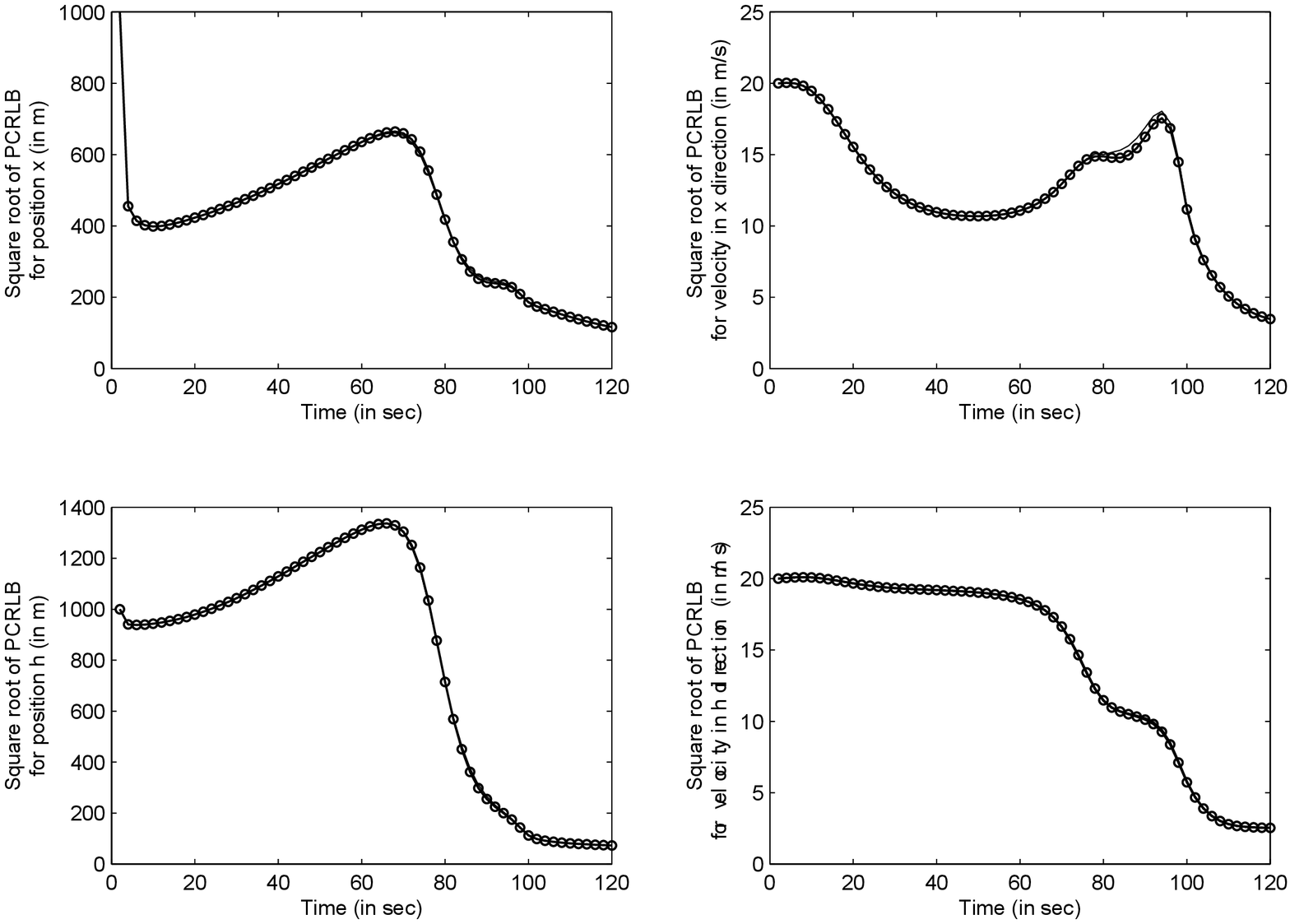}
        }%
        \subfigure[Square root of the theoretical (solid line with marker) and approximate PCRLB (solid line) for all the target states under Case 2]{%
          \label{fig:Chap2F1b}
           \includegraphics[scale=0.40]{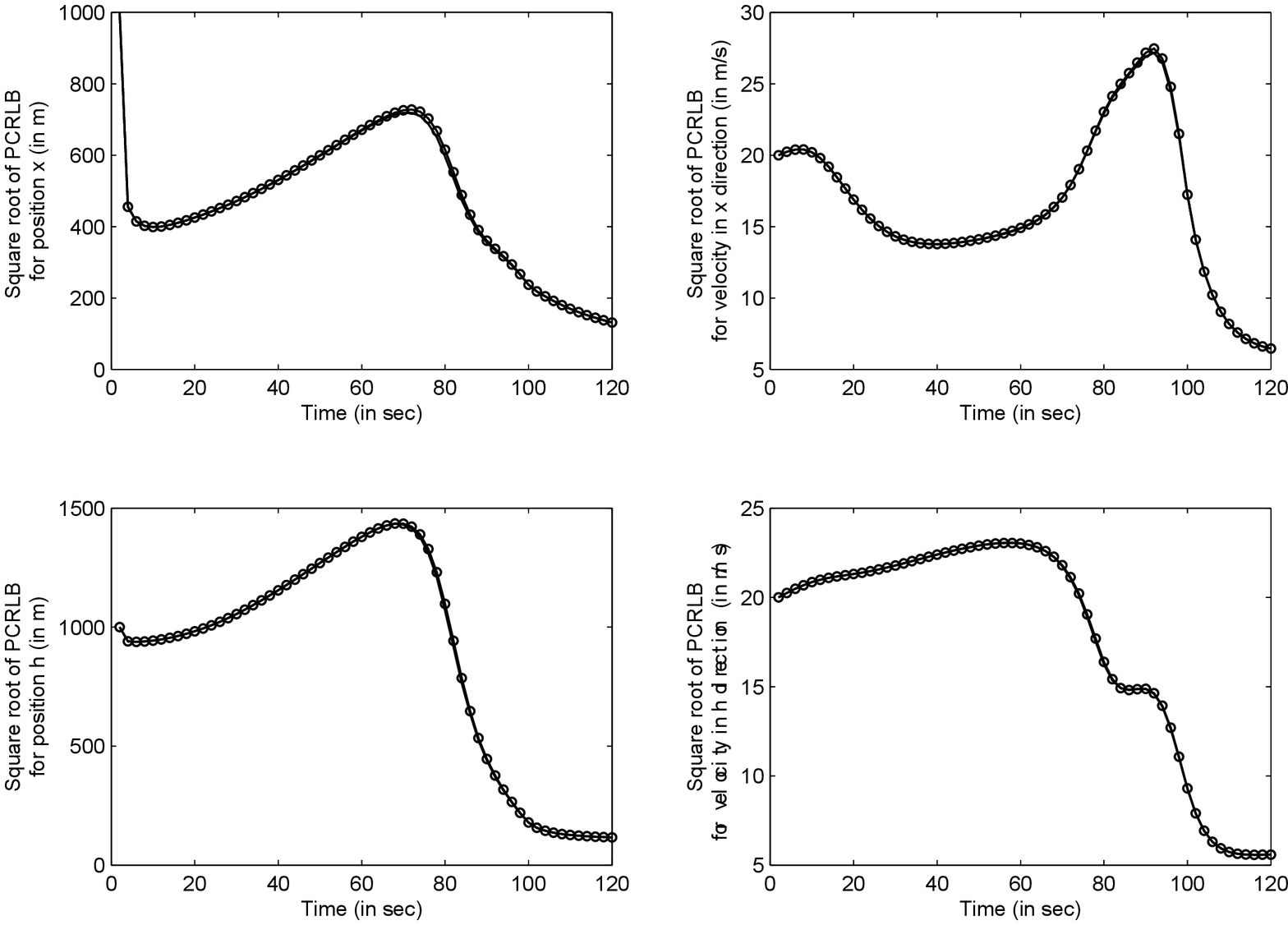}
        }\\ 
        \subfigure[Square root of the theoretical (solid line with marker) and approximate PCRLB (solid line) for all the target states under Case 3]{%
            \label{fig:Chap2F1c}
            \includegraphics[scale=0.40]{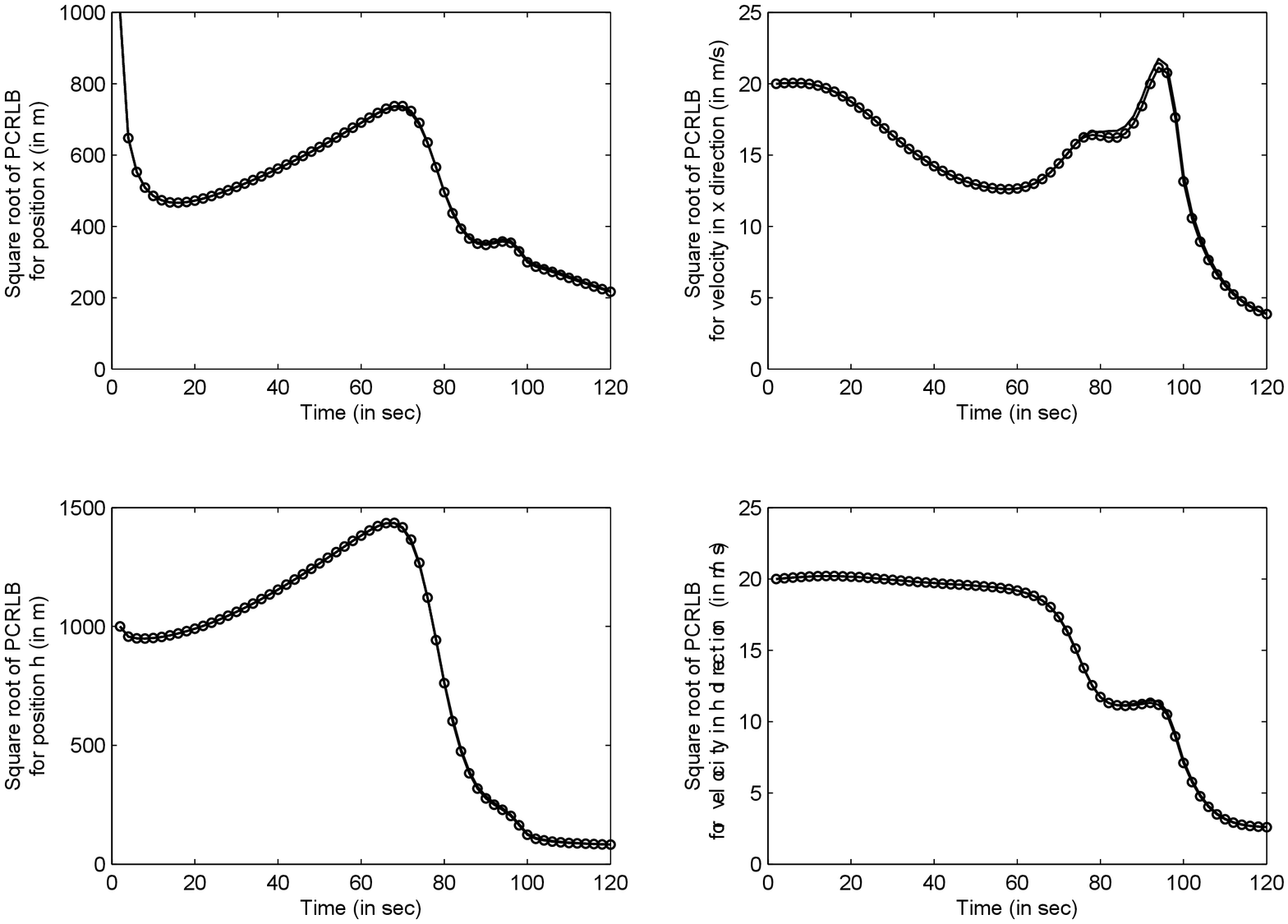}
        }%
        \subfigure[Square root of the theoretical (solid line with marker) and approximate PCRLB (solid line) for all the target states under Case 4]{%
            \label{fig:Chap2F1d}
            \includegraphics[scale=0.40]{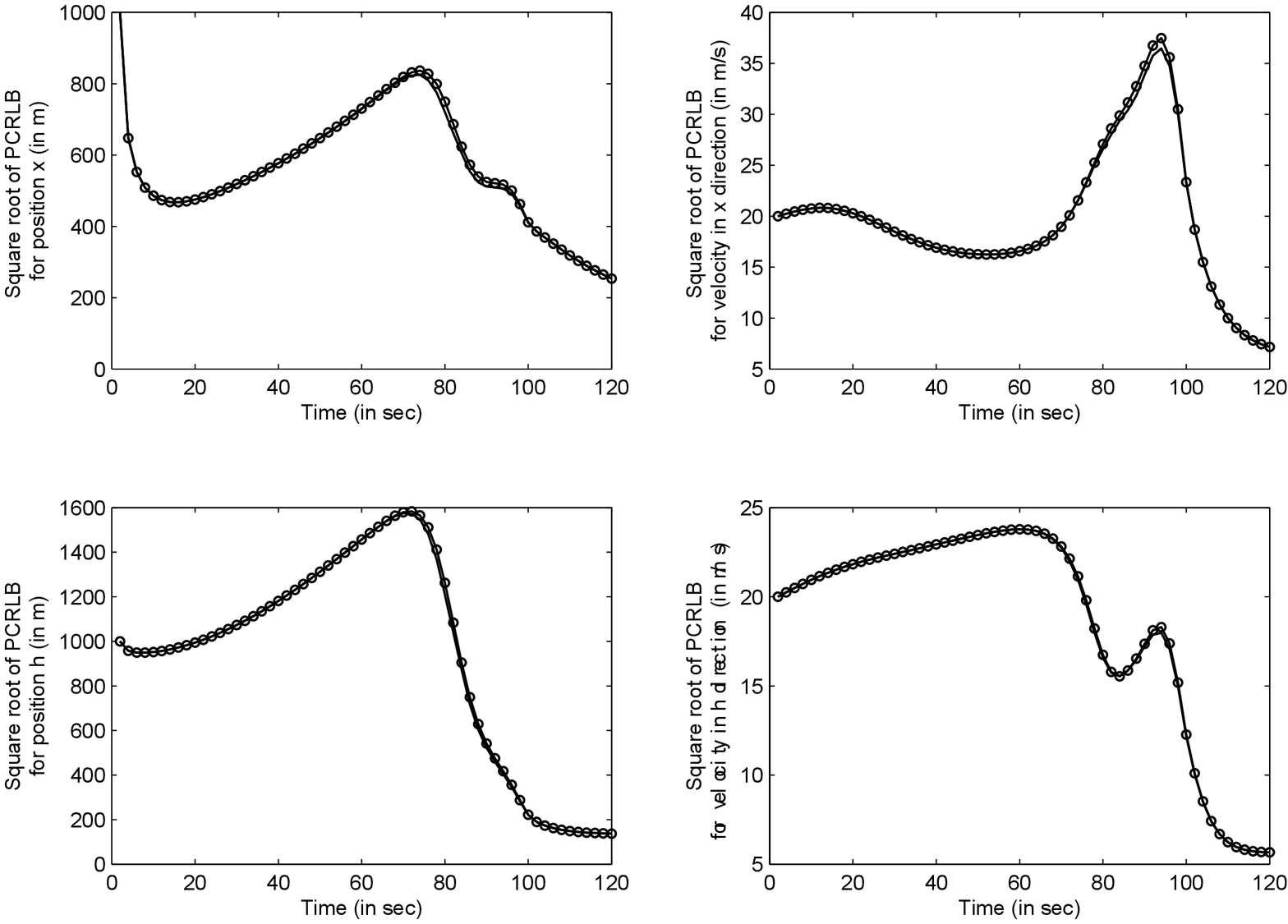}
        }\\ 
        \subfigure[Square root of the approximate PCRLBs for the target states under the cases listed in Table \ref{tab:Chap2Tab2}.]{%
            \label{fig:Chap2F6}   
            \includegraphics[scale=0.40]{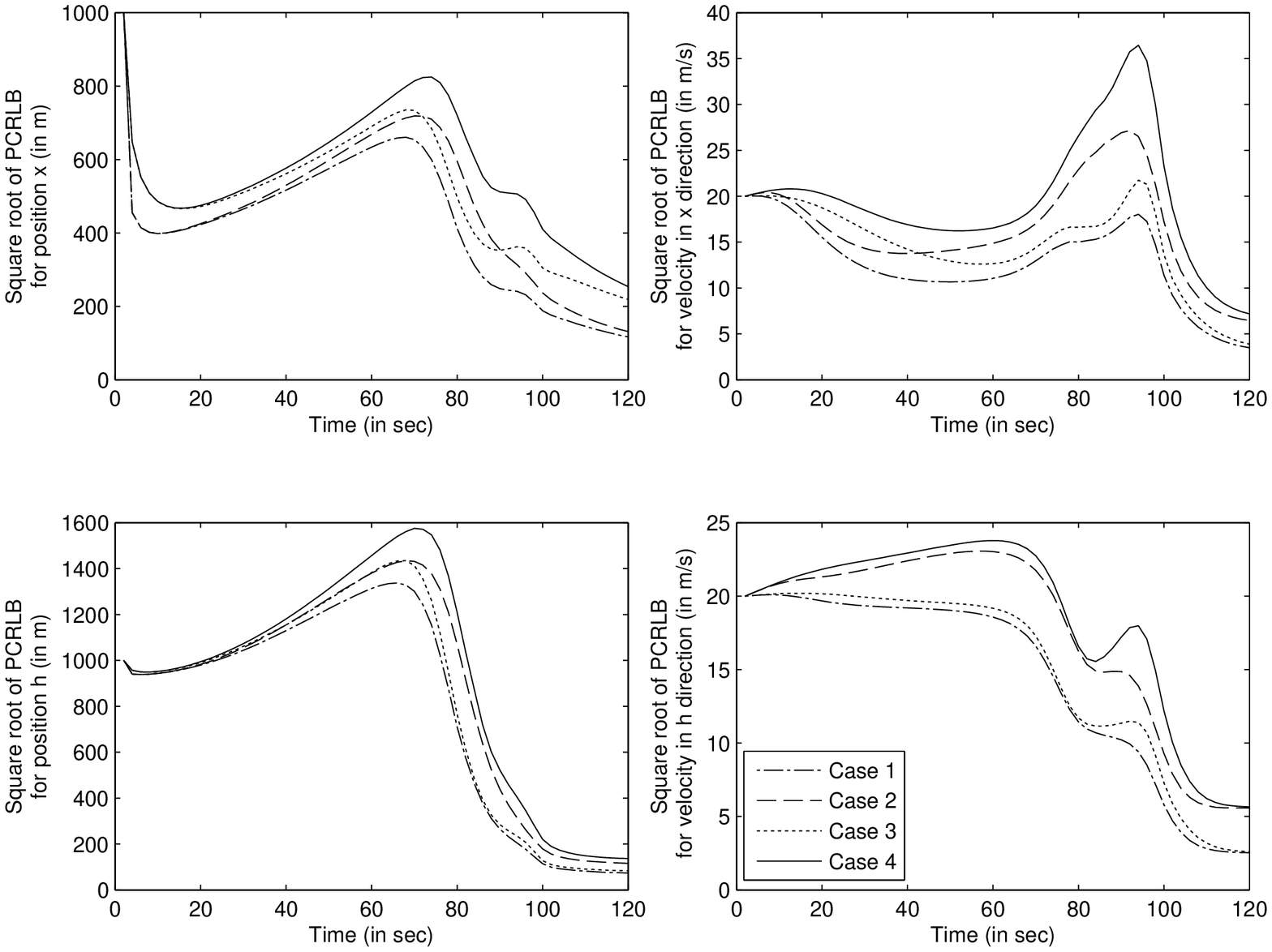}
        }%
        \subfigure[Square root of the theoretical and approximate PCRLBs for different values of $N$ in Example 1, Case 4. Note that all the sub-figures have been appropriately scaled up allow clear illustration of the effect of $N$ on the quality of approximation.]{%
            \label{fig:Chap2F7}
            \includegraphics[scale=0.40]{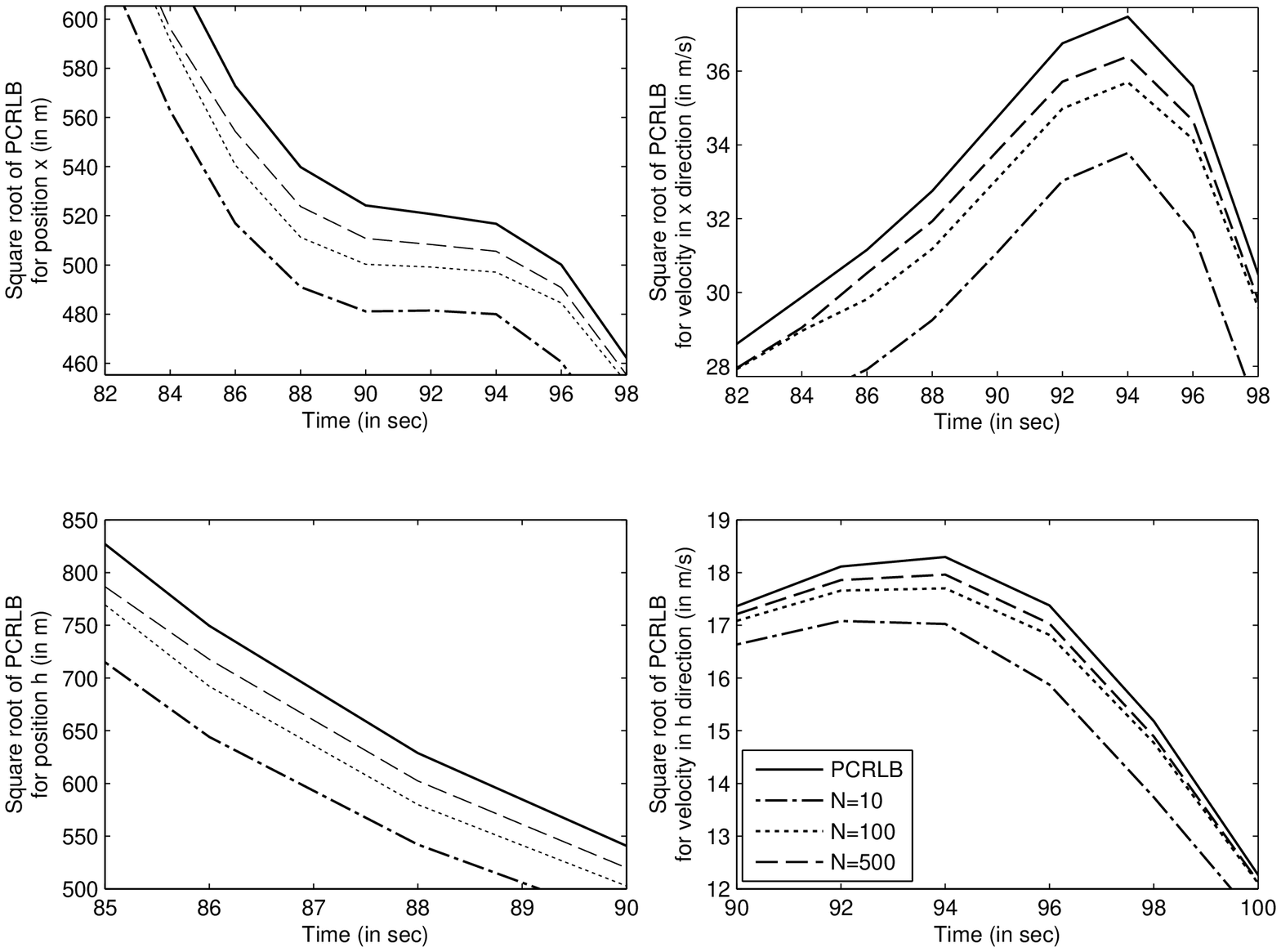}
        }%
    \end{center}
    \caption{%
        Results for Simulation Example 1.
     }%
\end{figure*}

\noindent \emph{Case 1:} Figure \ref{fig:Chap2F1} compares the square root of the SMC based approximate bound against the theoretical PCRLB. Clearly, the approximate bound for both the position and velocity of the target in both $\text{X}$ and $\text{H}$ coordinates accurately follows the theoretical bound at all tracking time instants. Note that the high values of the PCRLB in Figure\ref{fig:Chap2F1} highlights tracking difficulties as the target approaches the ground.\\

\emph{Case 2:} In this case the state noise intensity is increased five fold and the sensor noise is kept at a small value (see Table \ref{tab:Chap2Tab2}). Notwithstanding the increased noise variance, the PCRLB approximation is almost exact at all tracking time instants. The results for Case 2 are shown in Figure \ref{fig:Chap2F1b}. Table \ref{tab:Chap2Tab4} compares the $\Lambda_J$ values for Case 2 computed using (\ref{eq:Chap2Extra1}). Based on Table \ref{tab:Chap2Tab4}, the results from Cases 1 and 2 closely compare in terms of the order of the $\Lambda_J$ values. To allow further comparison with Case 1, the square root of the approximate PCRLBs for Cases 1 and 2 are compared in Figure \ref{fig:Chap2F6}. In terms of the magnitude, the PCRLB for Case 2 is higher than that for Case 1, suggesting tracking difficulties with larger noise intensity.\\
\emph{Case 3:} Again for Case 3, performance similar to Figure \ref{fig:Chap2F1} is obtained as given in Figure \ref{fig:Chap2F1c}. The same is evident from Table \ref{tab:Chap2Tab4}, where the average sum of square of error in approximating the PCRLB for Cases 1 and 3 are of the same order.\\
\emph{Case 4:} Results for Case 4 is given in Figure \ref{fig:Chap2F1d}. Higher values of the PCRLB for Case 4 in Figure \ref{fig:Chap2F6} reaffirms the estimation issues associated with larger noise variances. Similar conclusions can be drawn based on Table \ref{tab:Chap2Tab4}, where the $\Lambda_J$ values for Case 4 are the highest compared to the previous cases. Nevertheless, the errors are bounded and within a few orders of the $\Lambda_J$ values reported for Case 1.  

\begin{table}
\caption{Average sum of square of errors in approximating the PCRLB for the states in Example 1, under the cases in Table \protect\ref{tab:Chap2Tab2}.}
\centering
\begin{tabular}{lcccc}
\hline 
$\Lambda_J$ values& Case 1 & Case 2 & Case 3 & Case 4 \\ 
\hline 

$\Lambda_J(1,1)~(\times 10^{-6})$ & 9.30 & 50.7 & 5.87 & 130 \\ 
 
$\Lambda_J(2,2)~(\times 10^{-11})$ & 4.50 & 2.06 & 7.08 & 46.2 \\ 

$\Lambda_J(3,3)~(\times 10^{-5})$ & 3.56 & 23.1 & 2.96 & 100 \\ 

$\Lambda_J(4,4)~(\times 10^{-13})$ & 8.63 & 24.8 & 19.6 & 122 \\ 
\hline 
\end{tabular}
\label{tab:Chap2Tab4}
\end{table}
All the above case studies suggest that the proposed approach is accurate in approximating the theoretical PCRLB under large state and sensor noise variances.

\begin{rmk}
\label{remark:Chap2R10}
Note that in \cite{Lei2011}, a similar ballistic target tracking problem at re-entry phase was considered to illustrate the use of an EKF and UKF based method in approximating the theoretical PCRLB. Unlike the non-linear sensor model considered here (see (\ref{eq:Chap2E35})), \cite{Lei2011} used the change of coordinates method to obtain a linear sensor model representation. It is important to highlight that even with a linear sensor model, the EKF and UKF based method yields a biased estimate of the PCRLB for the target states (see Figures 4 through 7 in \cite{Lei2011}). Whereas, under a more challenging situation, as one considered here, the SMC based method yields an unbiased estimate of the PCRLB (see Figures \ref{fig:Chap2F1} through \ref{fig:Chap2F1d}, and Table \ref{tab:Chap2Tab4}). This highlights the advantages of the SMC based method (both in terms of the accuracy and applicability) over the EKF and UKF based PCRLB in presence of strong system or sensor non-linearities.
\end{rmk}

Next we study the sensitivity of the involved SMC approximations to the number of particles used. In Figure \ref{fig:Chap2F7}, approximate PCRLB bounds are compared against the theoretical PCRLB for different values of $N$. The results are obtained by varying $N$ in Algorithm 1. From Figure \ref{fig:Chap2F7}, it is clear that by simply increasing $N$, which is a tuning parameter in Algorithm 1, the quality of the SMC approximations can be significantly improved. For all the simulation cases, the number of Monte Carlo simulations was selected as ${M=200}$ (see Table \ref{tab:Chap2Tab3}). Computation of a single Monte Carlo simulation took 0.69 seconds on a 3.33 GHz Intel Core i5 processor running on Windows 7. Note that the reported absolute execution time is solely for instructive purposes and is not intended to reflect on the true computational complexity of the proposed algorithm. Collectively, from Figures \ref{fig:Chap2F1} through \ref{fig:Chap2F7}, it is evident that the SMC based method is accurate in approximating the theoretical PCRLB for a range of target state and sensor noise variances.
\subsection{Example 2: A non-linear and non-Gaussian system}
\label{sec:Chap2S7.2}
The aim of this study is to demonstrate the effectiveness of the proposed SMC based method in approximating the PCRLB in presence of a non-Gaussian noise. 
\subsubsection{Model setup}
A more challenging situation is considered in this section that involves the following discrete-time, uni-variate non-stationary growth model 
\begin{subequations}
\label{eq:Chap2E38}
\begin{align}
X_{t+1}=&\frac{X_t}{2}+\frac{25X_t}{1+X_t^2}+8\cos{(1.2t)}+V_t,\label{eq:Chap2E38a}\\
Y_t=&\frac{X_t^2}{20}+W_t \label{eq:Chap2E38b},
\end{align}
\end{subequations}
where ${V_t\in\mathbb{R}}$ is an i.i.d.~sequence following a Gaussian distribution, such that ${V_t\sim \mathcal{N}(v_t|0,Q_t)}$. The noise variance is defined as $Q_t=5\times 10^{-3}~\forall t\in[1,T]$, where $T$ is $30$ seconds. Also, the initial state is modelled as ${X_0\sim\mathcal{N}(x_0|0,0.01)}$. This example has been profiled due to it being acknowledged as a benchmark problem in non-linear state estimation in several previous studies \cite{D2001,HB2004}.     
\subsubsection{Simulation setup}
To compute the SMC based approximate PCRLB solution, two different sensor noise models are considered in (\ref{eq:Chap2E38b}). For Case 1, $W_t\in\mathbb{R}$ is an i.i.d.~sequence following a Gaussian distribution, such that $W_t\sim \mathcal{N}(w_t|0, R_t)$, while for Case 2, $W_t\in\mathbb{R}$ is again an i.i.d~sequence, but follows a Rayleigh distribution, such that $W_t\sim \mathcal{R}(w_t|R_t)$. For both the cases, the sensor noise variance ${R_t=1\times 10^{-3}~\forall t\in[1,T]}$ is considered. Here Case 2 represents a much more challenging situation, where estimation is considered under a non-Gaussian sensor noise. For fair comparison, ${M=200}$ and ${N=100}$ are selected.
\subsubsection{Results}
\begin{figure}[t!]
   \centering
   \includegraphics[scale=0.42]{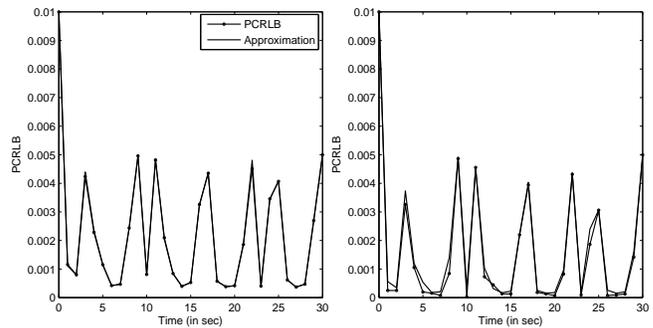}
   \caption{{Comparing the approximate PCRLB against the theoretical PCRLB in Example 2 under Gaussian (left) and Rayleigh (right) sensor noise distributions.}}
   \label{fig:Chap2F8}
\end{figure}

\noindent \emph{Case 1:} Comparison of the approximate and the theoretical PCRLB for the Gaussian sensor noise case is given in Figure \ref{fig:Chap2F8}. The results suggest that for the chosen $N$, the approximate PCRLB almost exactly follows the theoretical PCRLB at all filtering time instants. The same is reflected in the error value computed using (\ref{eq:Chap2Extra1}), which is $\Lambda_J=4.19\times 10^{-9}$. 

\noindent \emph{Case 2:} Figure \ref{fig:Chap2F8} compares the approximate PCRLB solution against the theoretical PCRLB for the Rayleigh sensor noise case. Although the approximation almost exactly follows the theoretical solution, compared to Case 1, the approximation is relatively coarser at certain time instants. This highlights the issues associated with estimation under non-Gaussian noise with limited $N$. Finally, the $\Lambda_J$ value for Case 2 is ${4.62\times 10^{-8}}$, which is within an order of the value reported for Case 1.

The simulation study clearly illustrates the efficacy of the proposed method in approximating the PCRLB for non-linear SSMs with non-Gaussian noise.
\section{Discussions}
\label{sec:Chap2S8}
The simulation results in Section \ref{sec:Chap2S7} demonstrate the utility and performance of the SMC based PCRLB approximation method developed in this paper. It is important to highlight that despite of the many convergence results discussed in Section \ref{sec:Chap2S6}, the choice of an SMC method plays a crucial role in determining the quality of the PCRLB approximation. Here, the use of a sequential-importance-resampling (SIR) filter of \cite{G2008,S2011} is motivated by the fact that it is relatively less sensitive to large state noise and is computationally less expensive. Furthermore, the importance weights are easily evaluated and the importance functions can be easily sampled \cite{R2004}; however, other algorithms such as Auxiliary-SIR (ASIR) \cite{APF1999} or Regularized PF (RPF) \cite{RPF2001} algorithm can also be used in place of SIR, as long as they are consistent with the approach developed herein.

An appropriate choice of the resampling method in Algorithm \ref{algorithm:Chap2A0}  is also crucial as it can substantially improve the quality of the approximations. The choice of the systematic resampling is supported by an easy implementation procedure and the low-order of computational complexity $\mathcal{O}(N)$ \cite{D2001}. Other resampling schemes such as stratified sampling \cite{K1996} and residual sampling \cite{LC1998} can also be used as an alternative to systematic resampling in the proposed framework.

In summary, with the aforementioned options, coupled with the user-defined choice of the parameters $N$ and $M$, an SMC based PCRLB approximation approach provides an efficient control over the numerical quality of the solution.
\section{Conclusions}
\label{sec:Chap2S9}
In this paper a numerical method to recursively approximate the PCRLB in \cite{T1998} for a general discrete-time, non-linear SSMs operating with ${\pr_d=1}$ and ${\pr_f=0}$ is presented. The presented method is effective in approximating the PCRLB, when the true states are hidden or unavailable. This has practical relevance in situations; wherein, the test-data consist of only sensor readings. The proposed approach makes use of the sensor readings to estimate the hidden true states, using an SMC method. The method is general and can be used to compute the lower bound for non-linear dynamical systems, with non-Gaussian state and sensor noise. The quality and utility of the SMC based PCRLB approximation was validated on two simulation examples, including a practical problem of ballistic target tracking at re-entry phase. The analysis of the numerical quality of the SMC based PCRLB approximation was investigated for a range of target state and sensor noise variances, and with different number of particles. The proposed method exhibited acceptable and consistent performance in all the simulations. Increasing the number of particles was in particular, found to be effective in reducing the errors in the PCRLB estimates. Finally, some of the strategies for improving the quality of the SMC based approximations were also discussed. 

The current paper assumes the model parameters to be known a priori; however,  for certain applications, this assumption might be a little restrictive. Future work will focus on extending the results of this work to handle such situations. Furthermore, use of SMC method in approximating the modified versions of the PCRLB, which allow tracking in situations, such as: target generated measurements; measurement origin uncertainty; cluttered environments; and Markovian models will also be considered.
\section*{Acknowledgement}
This work was supported by the Natural Sciences and Engineering Research Council (NSERC), Canada.
\bibliographystyle{IEEETran}
\bibliography{ifacconf}
\end{document}